\documentclass[a4paper,11pt]{article}
\usepackage{jheppub} 
\usepackage{lineno}

\usepackage{tikz}
\usetikzlibrary[snakes]
\usepackage{eufrak}
\usepackage{mathrsfs}
\usepackage{bbm}
\usepackage{mathabx}
\usepackage{tikz}
\usetikzlibrary[snakes]

\usepackage{graphicx}
\usepackage{subcaption}
\graphicspath{ {Images/} }

\title{\boldmath The linear property of genus-$g$, $n$-point, $b$-boundary, $c$-crosscap correlation functions in two-dimensional conformal field theory.}

\author{Xun Liu}
\affiliation{Department of Physics, The University of Tokyo \\
7-3-1 Hongo, Bunkyo-ku, Tokyo 113-0033, Japan}
\affiliation{RIKEN Interdisciplinary Theoretical and Mathematical Sciences (iTHEMS) \\  Wako, Saitama 351-0198, Japan}

\abstract{We propose a method to challenge the calculation of genus-$g$, bulk $n$-point, $b$-boundary, $c$-crosscap correlation functions with $x$ boundary operators $\mathcal{F}_{g,n,b,c}^{x}$ in two-dimensional conformal field theories (CFT$_2$). We show that $\mathcal{F}_{g,n,b,c}^{x}$ are infinite linear combinations of genus-$g$, bulk $(n+b+c)$-point functions $\mathcal{F}_{g,(n+b+c)}$, and try to obtain the linear coefficients in this work. We show the existence of a single pole structure in the linear coefficients at degenerate limits. A practical method to obtain the infinite linear coefficients is the free field realizations of Ishibashi states. We review the results in Virasoro minimal models $\mathcal{M}(p,p')$ and extend it to the $N=1$ minimal models $\mathcal{SM}(p,p')$.}

\begin{document}
\maketitle
\flushbottom

\ 

\section{Introduction}
\label{sec: intro}

The correlation functions are the main physical observable quantities of interest in quantum field theory. Among all known quantum field theories, two-dimensional conformal field theories (CFT$_2$) are a unique class, where the infinite-dimensional local conformal algebras provide much stronger analytical control of the theory \cite{Belavin:1984vu}. In principle, the correlation functions of CFT$_2$ on generic Riemann surfaces can be obtained from two distinct methods
\begin{itemize}
    \item  The solutions of conformal Ward identities on the surface of interest \cite{Belavin:1984vu}. 

    \item  The sewing of correlation functions on surfaces with lower dimensional moduli space, where a set of necessary constraint conditions needs to be satisfied \cite{Sonoda:1988mf, Sonoda:1988fq}. For rational CFT (RCFT), the constraint conditions include 
    \begin{itemize}
        \item The crossing symmetry of sphere four-point functions \cite{Belavin:1984vu}, which is equivalent to the associativity of bulk operators, 
        
        \item the modular covariance of torus correlation functions \cite{Cardy:1986ie, Sonoda:1988fq},

        \item the upper-half plane (UHP) sewing constraints \cite{Lewellen:1991tb, Runkel:1998he},

        \item the crosscap sewing constraints \cite{Fioravanti:1993hf}.
    \end{itemize}  
\end{itemize}
The calculations of correlation functions using the above two approaches are difficult and are limited to a few surfaces and theories. In this work, we propose a method to challenge this problem by utilizing the so-called linear property of the genus-$g$, $n$-point, $b$-boundary, $c$-crosscap functions with $x$ boundary operators $\mathcal{F}_{g,n,b,c}^{x}$.

Our method is valid for CFT$_2$ with state-operator correspondence, the concept of boundary and crosscap states. We also require the CFT$_2$ Hilbert spaces to take the form of 
\begin{equation}
    \mathcal{H} = \int_{ (i,\Bar{i}) }^{\oplus}    (  \mathcal{H}_{i} \otimes \Bar{\mathcal{H}}_{\Bar{i}} ) , 
\end{equation}
where $\mathcal{H}_{i}$ and $\Bar{\mathcal{H}}_{\Bar{i}}$ denote irreducible representations of chiral algebra $\mathcal{A}$ and anti-chiral algebra $\Bar{\mathcal{A}}$ respectively (the heterotic cases where $\mathcal{A} \ne \Bar{\mathcal{A}}$ included). For RCFTs, we will always consider the maximally extended theory, whose Hilbert spaces always take the form of \cite{Moore:1988ss, Moore:1988qv}
\begin{equation}
    \mathcal{H}= \bigoplus_{ i }  \: \mathcal{H}_{i} \otimes \mathcal{H}_{\lambda (i)} ,
\end{equation}
where $\lambda$ is a fusion rule automoprhism such that $N_{ij}^{k} =  N_{\lambda (i) \lambda(j) }^{ \lambda(k) }$ and $\lambda(0)=0$, where $N_{ij}^{k}$ are the fusion coefficients and $0$ denotes the vacuum module. Whether the torus partition functions are modular invariant is irrelevant to obtaining the linear coefficients. Hence, we don't assume the modular invariance in general.

To visualize the linear property, we express genus-$g$, bulk $n$-point, $b$-boundary, $c$-crosscap functions $\mathcal{F}_{g,n,b,c}^{x}$ correlation functions with $x$ boundary operators in general CFT$_2$ as  
\begin{equation*}
    \mathcal{F}_{g,n,b,c}^{x} [ \otimes_{i} \vert \phi_{\alpha_{i}} \rangle,  \otimes_{j} \vert (b_{ \{ j \} })_{\cdots} \rangle_{\{  \omega_{b_{j}} \} }^{r_{b_{j}}} , \otimes_{k} \vert  c_{k} \rangle_{\omega_{c_{k} }}^{r_{c_{k} }} ]  := 
\end{equation*}
\begin{equation}
   \langle  \Sigma_{g,(n+b+c)} \vert  \otimes_{i=1}^{n} \vert \phi_{\alpha_{i} } \rangle \otimes_{j=1}^{b} \vert (b_{ \{j \} })_{\cdots} \rangle_{\{  \omega_{b_{j}} \} }^{r_{b_{j}}}  \otimes_{k=1}^{c} \vert c_{k}\rangle^{r_{c_{k}}}_{\omega_{c_{k}}} \in \mathbb{C} ,\label{eq: def F gnbc}
\end{equation}
where $\langle \Sigma_{g,(n+b+c)} \vert$ is a surface state of a genus-$g$, $(n+b+c)$-boundary Riemann surface \cite{Alvarez-Gaume:1987eux, Sen:2014pia}. $\vert (b_{ \{j \} })_{\cdots} \rangle_{\{  \omega_{b_{j}} \} }^{r_{b_{j}}}$ and $\vert c_{k}\rangle^{r_{c_{k}}}_{\omega_{c_{k}}}$ are the defining boundary and crosscap states, where all information at the boundaries and crosscaps are encoded in. By expanding all the defining boundary states and crosscap states as infinite linear combinations of asymptotic states, all $\mathcal{F}_{g,n,b,c}^{x}$ can be written as infinite linear combinations of genus-$g$, bulk $(n+b+c)$-point bulk correlators $\mathcal{F}_{g,(n+b+c)}$. From which, the difficulties in the calculations of $\mathcal{F}_{g,n,b,c}^{x}$ are decomposed into
\begin{itemize}
    \item  Obtaining the linear coefficients of expanding the boundary states and crosscap states in asymptotic states, which is the equivalence of obtaining all infinite inversion Gram matrix elements, disk bulk one-point functions (with an arbitrary number of boundary operators), and crosscap one-point functions.

    \item  The calculations of all genus-$g$, $(n+b+c)$-point functions $\mathcal{F}_{g,(n+b+c)}$ in the expansion of $\mathcal{F}_{g,n,b,c}^{x}$.
\end{itemize}
The step of obtaining the linear coefficients can be further decomposed into three steps
\begin{itemize}
    \item  We remove all the boundary operators using consecutive applications of boundary-boundary OPEs and the inversion of bulk-boundary OPEs \cite{Cardy:1991tv, Lewellen:1991tb}. 

    \item  Boundary states without boundary operators and crosscap states are spanned boundary and crosscap Ishibashi states respectively \cite{Ishibashi:1988kg, Onogi:1988qk}. The linear coefficients are obtained by solving the Cardy's equation on an annulus \cite{Cardy:1989ir, Behrend:1998fd, Behrend:1999bn}, and its analogs on Mobius strip and Klein bottle   \cite{Pradisi:1995qy, Pradisi:1995pp}.

    \item  Finally, we need to expand all the Ishibashi states involved as infinite linear combinations of the asymptotic states, which is not practical in general.
\end{itemize}

The main focus of this work is to obtain the linear coefficients of expanding Ishibashi states in terms of asymptotic states. We attempt the direct method and show the existence of a single pole structure in the linear coefficients. We also show a more practical method: applying the free field resolution of chiral modules to Ishibashi states, since Ishibashi states in free field Fock spaces are coherent states \cite{Callan:1987px}.  We claim that a CFT$_2$ admits a free realization if it satisfies the following conditions \cite{Bouwknegt:1990wa}
\begin{itemize}
    \item  The realization of the chiral algebra $\mathcal{A}$ and antichiral algebra $\Bar{\mathcal{A}}$ in terms of the field fields.

    \item  The existence of projection maps from the free field Fock space modules to irreducible chiral representations in the CFT$_2$ Hilbert space, which is called the free field resolution. 

    \item The existence of a method to compute the correlation functions in the original CFT$_2$ using the free field vertex operators. 
\end{itemize}
The canonical example where all three requirements are satisfied is the Coulomb gas formalism of Virasoro minimal models $\mathcal{M}(p,p')$ (both unitary and non-unitary), where the Virasoro algebra with central charges $c_{p,p'}= 1- \frac{6(p-p')^{2}}{pp'}$ is realized by a background-charged boson \cite{Dotsenko:1984nm, Dotsenko:1984ad, Dotsenko:1985hi}. The resolution of irreducible chiral modules is realized by the zeroth cohomology spaces of some complexes of Fock space modules \cite{Felder:1988zp}. And the computation of the correlation functions of the Virasoro minimal model on spheres, surfaces of higher genus, and surfaces with boundaries can be computed from the vertex operators with extra insertion of screening operators \cite{Dotsenko:1984nm, Dotsenko:1984ad, Dotsenko:1985hi, Felder:1989ve, Kawai:2002pz}. If we only want to expand Ishibashi states in terms of asymptotic states, we may relax the third requirement, and consider more examples such as the free fermion expression. However, for our ultimate objective of calculating correlation functions $\mathcal{F}_{g,n,b,c}^{x}$, we should focus on the method satisfying all three requirements. The application of free field realization to Ishibashi states has been studied in several conventional studies with different motivations \cite{Kawai:2002vd, Kawai:2002pz, Parkhomenko:2003gy, Caldeira:2003zz, Hemming:2004dm, Parkhomenko:2004ab}. In this work, we review the application of Coulomb gas formalism to Virasoro minimal models and extend to $N=1$ minimal model cases. We also write down a free fermion expression of Ishibashi states in the Ising model. The application of free field realization to Ishibashi states in more CFT$_2$ is working in progress.

\ 

This work is organized as follows. In Section \ref{sec: def cor}, we describe the linear property of $\mathcal{F}_{g,n,b,c}^{x}$, and discuss the potential to express the UHP and crosscap sewing constraints purely in terms of bulk correlation functions. In Section \ref{sec: exp}, we attempt to obtain the linear coefficients by direct calculation and show a single pole structure. The single pole structure is tested at the lowest levels of Virasoro, $N=1$, $N=2$, and $\widehat{su(2)}_{k}$ Kac-Moody Ishbashi state expansions. In Section \ref{sec: fre ap}, we discuss the free field approach of Ishibashi states. We review the application of Coulomb gas formalism to Virasoro minimal models $\mathcal{M}(p,p')$ and extend to $N=1$ minimal $\mathcal{SM}(p,p')$ models. We also give the free fermion expression of Ising model Ishibashi states.

\ 

\section{The linear property of genus-$g$, bulk $n$-point, $b$-boundary, $c$-crosscap correlation functions $\mathcal{F}_{g,n,b,c}^{x}$ with $x$ boundary operators}
\label{sec: def cor}

\

In this section, we first review the concept of boundary and crosscap states. Next, we explain why correlation functions $\mathcal{F}_{g,n,b,c}^{x}$ can be written as (\ref{eq: def F gnbc}), discuss some results of this property, and the potential to express the UHP and crosscap sewing constraints in terms of purely bulk correlation functions.

\subsection{A quick review on boundary and crosscap states} \label{subsec: re bdy cro sta}

Consider a CFT$_2$ with conformal algebra $\mathcal{A} \otimes \Bar{\mathcal{A}}$, including the heterotic $\Bar{\mathcal{A}} \ne \mathcal{A}$ cases. We quickly review how to introduce the boundary and crosscap states, in which all the information on the corresponding boundaries and crosscaps is encoded.

Cardy discovered that the presence of boundaries reduces the $\mathfrak{Vir} \otimes \mathfrak{Vir}$ conformal algebra of Virasoro CFT$_2$ to an open-sector chiral Virasoro algebra $\mathfrak{Vir}^{op}$ \cite{Cardy:1984bb}. To show this, we consider a CFT$_2$ defined in the upper half-plane (UHP). The holomorphic and antiholomorphic energy-stress tensors have to be identical on the real axis $\mathbb{R}$ since energy flowing out of the boundary is prohibited 
\begin{equation}
    T(z) = \Bar{T}(\Bar{z}), \quad z= \Bar{z}.
\end{equation}
This condition allows us to view $\Bar{T}(\Bar{z})$ as the analytical continuation of $T(z)$ in the lower half-plane. Then, we can define a single set of Virasoro generators as 
\begin{equation}
    L_{n}^{UHP} := \int_{C} \frac{dz}{2\pi i} z^{n+1} T(z) - \int_{C} \frac{d\Bar{z}}{2\pi i} \Bar{z}^{n+1} \Bar{T}(z),  \label{eq: Vir op}
\end{equation}
which forms the open-sector chiral Virasoro algebra $\mathfrak{Vir}^{op}$. $C$ in (\ref{eq: Vir op}) is a semi-circle on the UHP whose center being $z=0$.

This technique can be generlized to generic conformal algebras $\mathcal{A} \otimes \Bar{\mathcal{A}} $ \cite{Recknagel:1997sb, Recknagel:2013uja}. Denote the maximal common subalgebra of $\mathcal{A}$ and $\Bar{\mathcal{A}}$ as $\mathcal{A}^{max}$, and its holomorphic generators as $\{ T(z), W^{i}(z) , \cdots  \}$. For a given boundary condition, some generators are preserved on the boundary \cite{Recknagel:1997sb, Recknagel:1998ih}, that is,
\begin{equation}
    T(z) = \Bar{T}(\Bar{z}), \quad W^{i} (z) = \Omega \Bar{W}^{i}(\Bar{z}), \quad z= \Bar{z}.  \label{eq: pre con bdry}
\end{equation}
The preserved generators form a chiral algebra $\mathcal{A}'$, and $\Omega$ is a point-wise automorphism of $\mathcal{A}'$, which acts trivially on the energy-stress tensor. The analytical continuation trick of energy-stress tensor (\ref{eq: Vir op}) applies to all the preserved generators
\begin{equation}
    (W_{n}^{i})^{op}:= \int_{C} \frac{dz}{2\pi i}  z^{n+h_{i}-1} W^{i}(z) - \int_{C} \frac{d\Bar{z}}{2\pi i}  \Bar{z}^{n+h_{i}-1}  \Omega \Bar{W}^{i}(\Bar{z}) .\label{eq: W op}
\end{equation}
Hence, the existence of the boundary breaks the original $\mathcal{A} \otimes \Bar{A}$ symmetry to a $( \mathcal{A}' )^{op}$ symmetry.

Next, we map the UHP to a unit disk and assign a boundary state to its boundary \cite{Recknagel:1997sb}. We expand the preserving conformal generators as Laurent modes and obtain the conditions for the boundary states
\begin{equation}
    (L_{n}- \Bar{L}_{-n}) \vert b \rangle = (W_{n}^{i}- \Omega (-1)^{h_{i}} \Bar{W}^{i}_{-n}) \vert b \rangle  =0 . \label{eq: Ishi con}
\end{equation}
The solutions of (\ref{eq: Ishi con}) is spanned by the ($\Omega$-twisted) boundary Ishibashi states $\vert  i ,b \rangle \rangle_{\omega} $ \cite{Ishibashi:1988kg, Onogi:1988qk}
\begin{equation}
    \vert i ,b \rangle \rangle_{\omega} = \sum_{N=0}^{\infty} \vert i,N \rangle \otimes  V_{\omega} U \vert i,N \rangle  \in \mathcal{H}_{i} \otimes \mathcal{H}_{\omega^{-1} (i^{\ast}) } \in \mathcal{H} ,
\end{equation}
where $\vert i,N \rangle$ denote the orthonormal basis of the irreducible module $\mathcal{H}_{i}$ of $\mathcal{A}'$. $U$ is an anti-unitary operator which maps $\mathcal{H}_{i}$ to its charge conjugate $\mathcal{H}_{i^{\ast}}$, and $V_{\omega}$ is an unitary isomorphism induced from $\Omega$ such that $V_{\omega} : \: \mathcal{H}_{i^{\ast}}  \to \mathcal{H}_{\omega^{-1}(i^{\ast})} $ \cite{Recknagel:2013uja}. 

Suppose that the boundary of the unit disk consists of boundary condition changing operators and multiple boundary conditions. In that case, we assign a boundary state $\vert (b_{\{ j \}})_{\cdots} \rangle_{ \{ \omega_{b_{j}} \} }  $ to encode all the information on the boundary. $\{ j \}$ denotes the set of boundary conditions on the $j^{\text{th}}$ boundary, $\{\omega_{b_{j}} \} $ is the set of gluing automorphism of $\{ j\}$, and $\cdots$ denotes the set of boundary operators. Boundary operators in CFT$_2$ can be introduced by two distinct methods
\begin{itemize}
    \item When a bulk operator $\phi_{(i,\Bar{i})}$ approaches a boundary with boundary condition $b_{1}$, boundary operators that don't change boundary conditions are given by the bulk-boundary OPEs \cite{Cardy:1991tv} 
\begin{equation}
    \phi_{(i,\Bar{i})}(z,\Bar{z})= \int_{k} C_{(i,\Bar{i})k}^{b_{1}}  \vert z-\Bar{z} \vert^{h_{k} - h_{i} -\Bar{h}_{\Bar{i}} } \psi_{k}^{b_{1}b_{1}} (x), \quad x= \frac{z+\Bar{z}}{2} ,  \quad \Bar{i} = \omega^{-1}_{1} (i^{\ast}) , \label{eq: buk-bdry OPE}
\end{equation}
where the integral runs over all irreducible representations $k$ in open-sector Hilbert space $\mathcal{H}^{b_{1} b_{1}}$. $C_{(i,\Bar{i})k}^{b_{1}}$ are the bulk-boundary OPE coefficients, which are non-zero if and only if $N_{i \omega^{-1}_{1} (i^{\ast}) }^{k} \ne 0$. In this work, we introduce the inversion of the bulk-boundary OPEs
\begin{equation}
    \psi_{i}^{b_{1} b_{1}} \big( \frac{z+\Bar{z}}{2} \big) = \int_{(a,\Bar{a})} \:  ^{ \text{in} }C_{(a,\Bar{a})b_{1}}^{i} \: \phi_{(a,\Bar{a}) }(z,\Bar{z}), \label{eq: inv bul-bdy ope}
\end{equation}
where the coefficients $^{ \text{in} }C_{(a,\Bar{a})b_{1}}^{i}$ cancels the dependence of $\vert z - \Bar{z} \vert$ on the RHS of (\ref{eq: inv bul-bdy ope}). The validity of the inversion bulk-boundary OPEs is unknown unless all the bulk-boundary OPE coefficients are known, which are obtained by solving the UHP constraint conditions \cite{Lewellen:1991tb, Runkel:1998he}. In some simple examples such as uncompactified free boson CFT$_2$, the inversion of bulk-boundary OPE is valid. 

\item Consider a finite cylinder with boundary conditions $b_{1}$ and $b_{2}$ at the two ends \cite{Cardy:1989ir}. The open-sector Hilbert $\mathcal{H}^{b_{1}b_{2}}$ can be decomposed into a direct integral over the chiral irreducible representations of the maximal common subalgebra of the preserved chiral algebras of $b_{1}$ and $b_{2}$
\begin{equation}
    \mathcal{H}^{b_{1}b_{2}} = \int_{k} \:  (\mathcal{H}_{k}^{b_{1}b_{2}} )^{ \oplus  n_{b_{1} b_{2}}^{k}   } , \quad  n_{b_{1} b_{2}}^{k} \in \mathbb{N} .  \label{eq: op-sec Hil decp}
\end{equation}
Cutting the finite cylinder open to a strip and extending its length to semi-infinite. By an exponential map from the semi-infinite strip to the UHP, we obtain a set of boundary condition changing operators at the origin $\psi_{k}^{b_{1}b_{2}} (0) $, each corresponding to an irreducible module in $\mathcal{H}^{b_{1} b_{2}}$. 
\end{itemize}

The OPE of a bulk operator approaching a crosscap only contains the identity operator \cite{Stanev:2001na}. Hence, all crosscap states don't carry operators. The constraint conditions for crosscap states are
\begin{equation}
    \big( L_{n}-(-1)^{n} \Bar{L}_{-n} \big) \vert c \rangle_{\omega}=0 , \quad  \big( W_{n}^{i}-(-1)^{n}(-1)^{h_{i}} \Bar{W}^{i}_{-n} \big) \vert c \rangle_{\omega}=0 .
\end{equation}
The solutions to these conditions are spanned by the crosscap Ishibashi states, which differ from the boundary Ishiabshi states by a phase factor
\begin{equation}
    \vert i ,c \rangle \rangle_{\omega} = e^{i\pi (L_{0}-h_{i}) } \vert i ,b \rangle \rangle_{\omega} 
\end{equation}

$\vert b,c \rangle^{r}$ appeared in (\ref{eq: def F gnbc}) are the so-called propagated boundary and crosscap states, defined as 
\begin{equation}
    \vert b,c \rangle^{r} = e^{  (L_{0} + \Bar{L}_{0}- \frac{c_{Vir}}{12} )  \ln ( r ) }  \vert b,c \rangle .
\end{equation}

\

\subsection{The linear property of $\mathcal{F}_{g,n,b,c}^{x}$} \label{subsec: def cor}

Now, consider the CFT$_2$ defined on a compact genus-$g$, $(n+b+c)$-boundary Riemann surface $\Sigma_{g,0,(n+b+c),0}$ (a compact genus-$g$, $n$-bulk punctured, $b$-boundary, $c$-crosscap Riemann surface with $x$ boundary punctures is denoted by $\Sigma_{g,n,b,c}^{x}$, $g,n,b,c ,x\in \mathbb{N}$). The complex local coordinates at each hole are denoted by $t_{i}$, $i=1,\cdots, (n+b+c)$, where the centers of the holes are set to be $t_{i}=0 $, and the boundaries are set to be the local unit circle $\vert t_{i} \vert =1$. The moduli space $\mathcal{M}_{g,0,(n+b+c),0}^{0}$ of the surface $\Sigma_{g,0,(n+b+c),0}^{0}$ has real dimension $[6g-6+3(n+b+c)]$, with $(n+b+c)$ real moduli are fixed to be $1$ by the radii of the holes. The path integral over the Riemann surface $\Sigma_{g,0,(n+b+c),0}^{0}$ gives a functional $\vert \Sigma_{g,(n+b+c)} \rangle$ depending on the boundary conditions at all the $(n+b+c)$ boundaries, indicating that the functional $\vert \Sigma_{g,(n+b+c)} \rangle $ is an element in the $(n+b+c)$-fold tensor product of the CFT$_2$ Hilbert space $\mathcal{H}$ \cite{Alvarez-Gaume:1987eux, Sen:2014pia}
\begin{equation}
   \vert \Sigma_{g,(n+b+c)} \rangle \in \mathcal{H}^{\otimes (n+b+c)}. 
\end{equation}
The dual element of $ \vert \Sigma_{g,(n+b+c)} \rangle$ is called a surface state of the surface $\Sigma_{g,0,(n+b+c),0}^{0}$
\begin{equation}
     \langle  \Sigma_{g,(n+b+c)} \vert \in (\mathcal{H}^{\ast} )^{\otimes (n+b+c)}  ,
\end{equation}
where $\mathcal{H}^{\ast}$ is the dual space of $\mathcal{H}$. The surface states consist of the geometric information of the $[6g-6+2(n+b+c)]$ unfixed real moduli of the surface $\Sigma_{g,0,(n+b+c),0}^{0}$. Hence, every surface state $ \langle  \Sigma_{g,(n+b+c)} \vert$ corresponds to a point in the moduli space of genus-$g$, $(n+b+c)$-punctured surface $\mathcal{M}_{g,(n+b+c)}$ and those points fully cover the moduli space $\mathcal{M}_{g,(n+b+c)}$.

\begin{figure}
    \centering
    \begin{tikzpicture}
  \shade[ball color = gray] (0,0) circle (2cm);
  \draw (0,0) circle (2cm);
  \fill[fill=white] (0.9,0.9) circle (0.6cm);
  \fill[fill=white] (-0.9,-0.1) circle (0.6cm);
  \fill[fill=white] (0.7,-0.7) circle (0.6cm);
 \draw (0,2.6) node {$\langle \Sigma_{0,3} \vert \in (\mathcal{H}^{\ast})^{\otimes 3} $}   ;  

\shade[ball color = gray] (6,0) circle (2cm);
  \draw (6,0) circle (2cm);
  \fill[fill=white] (5.1,-0.1) circle (0.4cm);
  \fill[fill=white] (6.7,-0.7) circle (0.4cm);
 \draw [thick, yellow] (6.75,0.75)--(7.05,1.05) ;
  \draw [thick, yellow] (6.75,1.05)--(7.05,0.75) ;
 \draw [red, thick] (5.1,-0.1) circle (0.41cm) ; 
 \draw [green, thick] (6.7,-0.7) circle (0.41cm) ; 
  \draw [green, thick] (6.4, -1.0) -- (7.0, -0.4) ; 

  \draw [green, thick] (6.4, -0.4) -- (7.0, -1.0) ; 

 \draw [red, thick] (4.6, -0.0) -- (4.8, -0.2) ; 

  \draw [red, thick] (4.6, -0.2) -- (4.8, -0.0) ; 

   \draw [red, thick] (5.4, -0.0) -- (5.6, -0.2) ; 

  \draw [red, thick] (5.4, -0.2) -- (5.6, -0.0) ; 

  \draw (3,0) node {$\Longrightarrow$}   ;  

\draw (6,2.6) node {$ \mathcal{F}_{0,1,1,1}^{2} = \langle \Sigma_{0,3}\vert \phi_{1} \rangle \otimes  \vert (b_{ \{2 \} })_{\psi_{1}\psi_{2}} \rangle_{\{  \omega_{b_{2}} \} }^{r_{b_{2}}} \otimes  \vert c_{3} \rangle_{  \omega_{c_{3}} }^{r_{c_{3}}} \in \mathbb{C}$ }   ;  

\end{tikzpicture}
    \caption{The construction of a sphere bulk one-point, one-boundary, and one-crosscap function with $2$ boundary operators $\mathcal{F}_{0,1,1,1}^{2}$.}
    \label{fig: 3 hole S^2}
\end{figure}
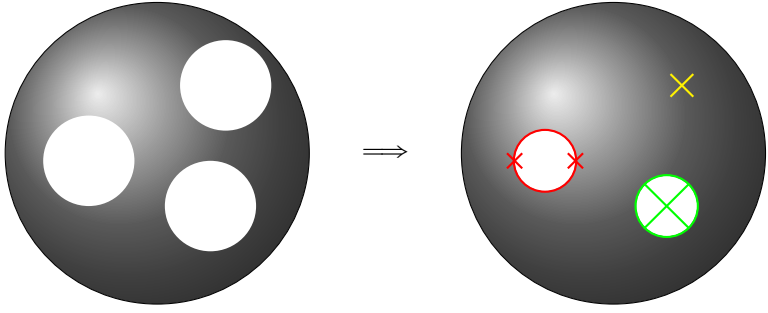

Genus-$g$, $n$-point, $b$-boundary, $c$-crosscap correlation functions $\mathcal{F}_{g,n,b,c}^{x} $ with $x$ boundary operators can be written as
\begin{equation*}
    \mathcal{F}_{g,n,b,c}^{x} [ \otimes_{i} \vert \phi_{\alpha_{i}} \rangle,  \otimes_{j} \vert (b_{ \{ j \} })_{\cdots} \rangle_{\{  \omega_{b_{j}} \} }^{r_{b_{j}}} , \otimes_{k} \vert  c_{k} \rangle_{\omega_{c_{k} }}^{r_{c_{k} }} ]  
\end{equation*}
\begin{equation}
    = \langle  \Sigma_{g,(n+b+c)} \vert  \otimes_{i=1}^{n} \vert \phi_{\alpha_{i} } \rangle \otimes_{j=1}^{b} \vert (b_{ \{j \} })_{\cdots} \rangle_{\{  \omega_{b_{j}} \} }^{r_{b_{j}}}  \otimes_{k=1}^{c} \vert c_{k}\rangle^{r_{c_{k}}}_{\omega_{c_{k}}} \in \mathbb{C} . \label{eq: def F gnbc 2}
\end{equation}
Geometrically, taking the inner product (\ref{eq: def F gnbc 2}) partially fill the $(n+b+c)$ holes on the original surface $\Sigma_{g,0,(n+b+c),0}^{0}$ by $n$ local disks with a single bulk puncture at the center $t_{i}=0$, $b$ local annulus with their inner boundaries at $\vert t_{j} \vert =r_{b_{j}}$ (having a total $x$ marked points on all the $b$ inner boundaries), and $c$ local crosscaps at regions $\vert t_{k} \vert \le  r_{c_{k}} $. The resulting surface  $\Sigma_{g,n,b,c}^{x}$ is a genus-$g$, $n$-bulk punctured, $b$-boundary, $c$-crosscaps Riemann surface with $x$ boundary punctures. A simple example of such a procedure is shown in Figure \ref{fig: 3 hole S^2}.

The correlation functions $\mathcal{F}_{g,n,b,c}^{x}$ depend on $(6g-6+2n+3b+3c+x)$ real moduli, where $(6g-6+2n+2b+2c)$ are from the surface state, $(b+c)$ are from the radii of boundaries and crosscaps, and $x$ are from the positions of the boundary punctures. The surfaces obtained from the above filling procedure live in the following region $\mathcal{R}$ of the moduli space $\mathcal{M}_{g,n,b,c}^{x}$ 
\begin{equation}
    \mathcal{R} \subset   \big\{ \mathcal{M}_{g,(n+b+c)} \times (0,1)^{(b+c)} \times [0,2\pi )^{x} \big\}  \subset \mathcal{M}_{g,n,b,c}^{x} , \label{eq: mod sp gnbc}
\end{equation}
where the boundary operators on the same boundary cannot be placed in the same position. We are uncertain about whether this region $\mathcal{R}$ can fully cover the moduli space $\mathcal{M}_{g,n,b,c}^{x}$. Nevertheless, we claim that the expression of the correlation functions $\mathcal{F}_{g,n,b,c}^{x}$ are valid over the whole moduli space $\mathcal{M}_{g,n,b,c}^{x}$. The reason is that the correlation functions $\mathcal{F}_{g,n,b,c}^{x}$ are solutions to Ward identities of the conformal algebra. The same correlation function $\mathcal{F}_{g,n,b,c}^{x}$ defined on surfaces at different points of the moduli space $\mathcal{M}_{g,n,b,c}^{x}$ should satisfy the same set of conformal Ward identities with different values of modular parameters, which indicates that the same correlation function $\mathcal{F}_{g,n,b,c}^{x}$ have the same expression at different points of the moduli space.

\paragraph{The linear property:} 

The fact that boundary and crosscap states are both elements of the CFT$_2$ Hilbert space $\mathcal{H}$ implies that correlation functions $\mathcal{F}_{g,n,b,c}^{x}$ are infinite linear combinations of genus-$g$, $(n+b+c)$-point functions $\mathcal{F}_{g,(n+b+c)}$. The general expression of linear coefficients of this expansion is obtained by inserting $(b+c)$ identity operators $I \otimes \Bar{I}$ expressed by the asymptotic states and their dual, denoted by $\vert \phi_{\alpha} \rangle$ and $ \langle (\phi_{\beta})^{c} \vert $ respectively
\begin{equation}
    I \otimes \Bar{I} = \int_{\forall \alpha , \beta \in \mathcal{H} }  \vert \phi_{\alpha}\rangle \:  (G)^{\alpha \beta}  \: \langle (\phi_{\beta})^{c} \vert , \quad  (G)_{\alpha \beta} : = \langle (\phi_{\beta} )^{c} \vert \phi_{\alpha} \rangle. \label{eq: I comp bas}
\end{equation}
Inserting this expression before all boundary and crosscap states in $\mathcal{F}_{g,n,b,c}^{x}$ (\ref{eq: I comp bas}) expand $\mathcal{F}_{g,n,b,c}^{x}$ as infinite linear combinations of $\mathcal{F}_{g,(n+b+c)}$. The expansion of a single boundary or crosscap state can be expressed as
\begin{equation}
   \vert b,c \rangle^{r} =   \int_{\alpha , \beta}  \:  \vert \phi_{\alpha} \rangle  (G)^{\alpha \beta}  \langle  (\phi_{\beta})^{c}  \vert b,c \rangle^{r} ,\label{eq; b c sta exp}
\end{equation}
where the linear coefficients consist of elements in the inversion Gram matrix $(G)^{\alpha \beta}$ and the inner products $\langle  (\phi_{\beta})^{c}  \vert b,c \rangle^{r}$, which are disk or crosscap bulk one-point functions.

The calculations of $(G)^{\alpha \beta}$ and $\langle  (\phi_{\beta})^{c}  \vert b,c \rangle^{r}$ are complicated and unpractical. The difficulties are mainly from two aspects
\begin{itemize}
    \item  There are infinite vectors in the Hilbert space $\mathcal{H}$ and the commutation relations of $\mathcal{A}$ are not simple in general. 

    \item   We need to know the explicit form of boundary and crosscap states. 
\end{itemize}

First, we attempt to remove the boundary operators. Combining the inversion bulk-boundary OPE (\ref{eq: inv bul-bdy ope}) with the boundary-boundary OPEs 
\begin{equation}
    \psi_{i_{j}}^{b_{j} b_{j+1} } (x_{j})\psi_{i_{j+1}}^{b_{j+1} b_{j+2} }(x_{j+1}) = \int_{ i_{a_{j}} } C_{ i_{j} i_{j+1} i_{a_{j}}  }^{ b_{j} b_{j+1} b_{j+2} }  \vert x_{j(j+1)}\vert  \psi_{i_{a_{j}}}^{b_{j} b_{j+2} } , \quad x_{j(j+1)}:= x_{j}-x_{j+1} ,
\end{equation}
we can remove the boundary operators in any $\mathcal{F}_{g,n,b,c}^{k}$, and express them as infinite linear combinations of bulk correlation functions. After removing all the boundary operators, there are three steps remain to compute the $\mathcal{F}_{g,n,b,c}^{0}$ in the expansion 
\begin{itemize}
    \item  Writing all the boundary and crosscap states as linear combinations of boundary Ishibashi and crosscap Ishibashi states.

    \item  Expanding all the Ishibashi states as infinite linear combinations of asymptotic states corresponding to bulk operators. 

    \item  Calculating all bulk correlation functions in the expansion. 
\end{itemize}
All the three steps are difficult, and we focus only on the second step in this work.

\

\subsection{The potential to express UHP and crosscap sewing constraints using bulk correlation functions \label{subsec: cons cond}} 

We discuss the consistency condition of correlation functions $\mathcal{F}_{g,n,b,c}^{x}$ using the linear property. Conventional studies have proved that the complete set of consistency conditions of CFT correlation functions includes the following 
\begin{itemize}
    \item The bulk OPE associativity \cite{Belavin:1984vu}. For some models, the torus partition function must be modular invariant \cite{ Cardy:1986ie,  Sonoda:1988fq}. 

    \item The UHP sewing constraints \cite{Lewellen:1991tb, Runkel:1998he}. 

    \item The crosscap sewing constraints \cite{Fioravanti:1993hf}. 
\end{itemize}

\begin{figure}[htbp]
    \centering
    \begin{tikzpicture}
      \draw [thick, black] (-6,0)--(-2,0);
      \draw [thick, black] (-5,0)--(-5,1);
      \draw [thick, black] (-4,0)--(-4,1);
      \draw [thick, black] (-3,0)--(-3,1);
      \draw [blue] (-5,1.25) node{$\vert b_{1} \rangle$} ; 
      \draw [green] (-4,1.25) node{$\vert b_{2} \rangle$} ; 
      \draw [red] (-3,1.25) node{$\vert b_{3} \rangle$} ; 
      \draw [thick, black, ->] (-2,1) -- (-1,2) ; 
      \draw [black] (-1.5,2) node{$F$} ;
      \draw [thick, black] (-1,3)--(2,3);
      \draw [thick, black] (0,3)--(0,4);
      \draw [thick, black] (1,3)--(1,3.5);
      \draw [thick, black] (1,3.5)--(2,3.5);
      \draw [thick, black] (1.5,3.5)--(1.5,4); 
      \draw [blue] (0,4.25) node{$\vert b_{1} \rangle$} ;
       \draw [green] (1.5,4.25) node{$\vert b_{2} \rangle$} ; 
      \draw [red] (2,3.75) node{$\vert b_{3} \rangle$} ; 
      \draw [thick, black, ->] (-2,1) -- (-1,2) ;  
      \draw [thick, black, ->] (3,3) -- (4,3) ;   
       \draw [black] (3.5,3.5) node{$F$} ;
       \draw [thick, black] (5,3)--(8,3);
      \draw [thick, black] (6,3)--(6,3.5);
      \draw [thick, black] (6,3.5)--(8,3.5);
      \draw [thick, black] (6.5,3.5)--(6.5,4);
      \draw [thick, black] (7.5,3.5)--(7.5,4);
      \draw [blue] (6.5,4.25) node{$\vert b_{1} \rangle$} ;
       \draw [green] (7.5,4.25) node{$\vert b_{2} \rangle$} ; 
      \draw [red] (8,3.75) node{$\vert b_{3} \rangle$} ; 
      \draw [thick, black, ->] (-2,1) -- (-1,2) ;  
      \draw [thick, black, ->] (3,3) -- (4,3) ;   
      \draw [thick, black, ->] (-2,-1) -- (-1,-2) ;
       \draw [thick, black] (-1,-3)--(2,-3);
      \draw [thick, black] (0,-3)--(0,-2.5);
      \draw [thick, black] (0,-2.5)--(1,-2.5);
      \draw [thick, black] (0.5,-2.5)--(0.5,-2);
      \draw [thick, black] (1.5,-3)--(1.5,-2); 
      \draw [blue]  (0.5,-1.75) node{$\vert b_{1} \rangle$} ;
       \draw [green] (1,-2.25) node{$\vert b_{2} \rangle$} ; 
      \draw [red] (1.5,-1.75) node{$\vert b_{3} \rangle$} ; 
       \draw [thick, black, ->] (3,-3) -- (4,-3) ;   
    \draw [black] (3.5,-2.5) node{$F$} ;
     \draw [black] (-1.5,-1) node{$F$} ;  
      \draw [thick, black] (5,-3)--(8,-3);
      \draw [thick, black] (6,-3)--(6,-2.5);
      \draw [thick, black] (6,-2.5)--(7,-2.5);
      \draw [thick, black] (6.5,-2.5)--(6.5,-2);
      \draw [thick, black] (6.5,-2)--(7.5,-2); 
      \draw [thick, black] (7,-2)--(7,-1.5); 
      \draw [blue]  (0.5,-1.75) node{$\vert b_{1} \rangle$} ;
       \draw [green] (1,-2.25) node{$\vert b_{2} \rangle$} ; 
      \draw [red] (1.5,-1.75) node{$\vert b_{3} \rangle$} ; 
      \draw [thick, black, ->] (6.5,0.5) -- (6.5,-0.5) ;
      \draw (7,0) node{$F$} ; 
      \draw [blue]  (7,-1.25) node{$\vert b_{1} \rangle$} ;
       \draw [green] (7.75,-2) node{$\vert b_{2} \rangle$} ; 
      \draw [red] (7.25,-2.5) node{$\vert b_{3} \rangle$} ; 
    \end{tikzpicture}
    \caption{The pentagon identity for sphere five-boundary functions $\mathcal{F}_{0,0,5,0}^{x}$, where the two endpoints without labeling also correspond to boundary states.}
    \label{fig: pen id sp 5 bdy}
\end{figure}
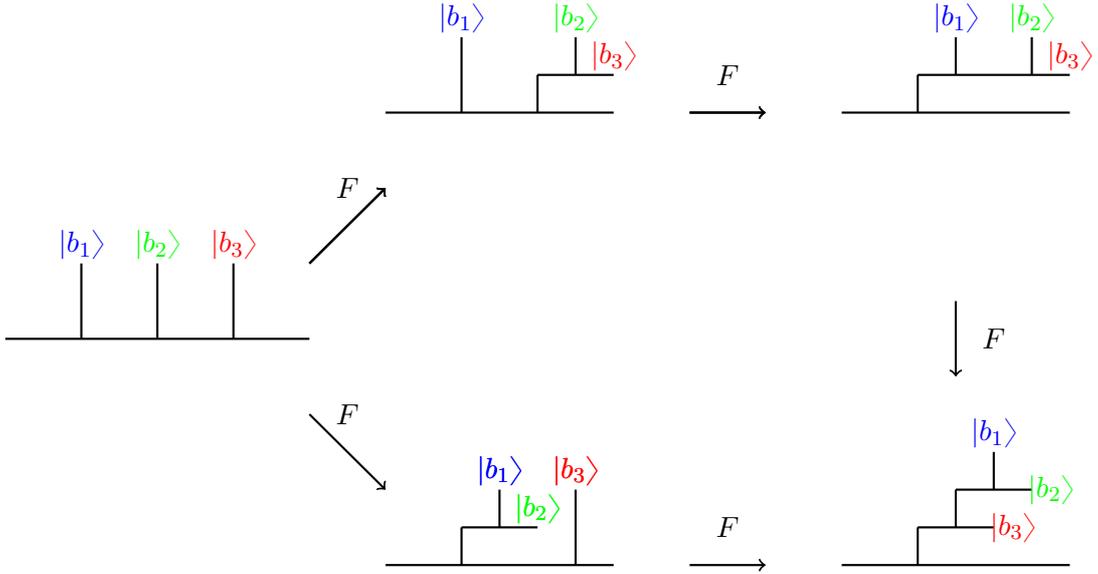

Consider we expand a correlation function $\mathcal{F}_{g,n,b,c}^{x}$ as infinite linear combinations of $\mathcal{F}_{g,(n+b+c)}$. The OPE associativity of all bulk correlation functions in the expansion is combined into the product associativity between bulk operators, boundary states, and crosscap states. We show a simple example in Figure \ref{fig: asso con}. This property indicates that $\mathcal{F}_{g,n,b,c}^{x}$ will satisfy duality identities similar to those of bulk correlation functions $\mathcal{F}_{g,(n+b+c)}$ \cite{Moore:1988qv}. As an example, we draw a pentagon identity for sphere five-boundary functions $\mathcal{F}_{0,0,5,0}^{x}$ in Figure \ref{fig: pen id sp 5 bdy}, where the fusing matrix $F$ acts on all the sphere five-point functions $\mathcal{F}_{0,5}$ in the expansion. 
\begin{figure}[htbp]
        \centering
        \begin{tikzpicture}
        \draw (-6,0) node {$\int$} ;
            \draw [black, thick] (-5,0)--(-4,0) ; 
            \draw [black, thick] (-5,0)--(-5.5,0.5) ; 
            \draw [black, thick] (-5,0)--(-5.5,-0.5) ; 
            \draw [black, thick] (-4,0)--(-3.5,0.5) ; 
            \draw [black, thick] (-4,0)--(-3.5,-0.5) ; 
            \draw [blue] (-5.5,0.8) node {$\vert b_{1} \rangle$} ;  
            \draw [purple] (-5.5,-0.8) node {$\vert b_{2} \rangle$} ;

            \draw [red] (-3.5,0.8) node {$\vert b_{3} \rangle$} ;  
            \draw [gray] (-3.5,-0.8) node {$\vert b_{4} \rangle$} ;

            \draw (-3,0) node {$=$} ;  

            \draw (-2.5,0) node {$\int$} ;

            \draw [black, thick] (-1.5,0)--(-0.5,0) ; 
            \draw [black, thick] (-1.5,0)--(-2,0.5) ; 
            \draw [black, thick] (-1.5,0)--(-2,-0.5) ; 
            \draw [black, thick] (-0.5,0)--(-0,0.5) ; 
            \draw [black, thick] (-0.5,0)--(-0,-0.5) ; 
             \draw [blue] (-1.5,0.8) node {$ \int c_{\alpha \beta 1} \vert \phi_{\alpha_{1}} \rangle $} ;  
            \draw [purple] (-1.5,-0.8) node {$ \int c_{\alpha \beta 2} \vert \phi_{\alpha 2} \rangle$} ;

            \draw [red] (0.5,0.8) node {$\int c_{\alpha \beta 3} \vert \phi_{\alpha 3} \rangle$} ;  
            \draw [gray] (0.5,-0.8) node {$\int c_{\alpha \beta 4} \vert \phi_{\alpha 4} \rangle$} ;

            \draw (0.5,0) node {$=$} ; 

            \draw (1.5,0) node {$\int$} ;

            \draw [black, thick] (2.5,0.5)--(2.5,-0.5)    ;

            \draw [black, thick] (2.5,0.5)--(2,1)    ; 

            \draw [black, thick] (2.5,-0.5)--(2,-1)    ;

            \draw [black, thick] (2.5,0.5)--(3,1)    ; 

            \draw [black, thick] (2.5,-0.5)--(3,-1)    ; 

             \draw [blue] (1.5,1.5) node {$ \int c_{\alpha \beta 1} \vert \phi_{\alpha 1} \rangle $} ;  
            \draw [purple] (1.5,-1.5) node {$ \int c_{\alpha \beta 2} \vert \phi_{\alpha 2} \rangle$} ;

            \draw [red] (3.5,1.5) node {$\int c_{\alpha \beta 3} \vert \phi_{\alpha 3} \rangle$} ;  
            \draw [gray] (3.5,-1.5) node {$\int c_{\alpha \beta 4} \vert \phi_{\alpha 4} \rangle$} ;
            \draw (4,0) node {$=$} ; 

            \draw (4.5,0) node {$\int$} ;

             \draw [black, thick] (5.5,0.5)--(5.5,-0.5)    ;

            \draw [black, thick] (5.5,0.5)--(5,1)    ; 

            \draw [black, thick] (5.5,-0.5)--(5,-1)    ;

            \draw [black, thick] (5.5,0.5)--(6,1)    ; 

            \draw [black, thick] (5.5,-0.5)--(6,-1)    ; 

            \draw [blue] (5,1.5) node {$ \vert b_{1} \rangle $} ;  
            \draw [purple] (5,-1.5) node {$ \vert b_{2} \rangle$} ;

            \draw [red] (6,1.5) node {$\vert b_{3} \rangle$} ;  
            \draw [gray] (6,-1.5) node {$\vert b_{4} \rangle$} ;

        \end{tikzpicture}
        \caption{The bulk OPE associativity of a $\mathcal{F}_{0,0,4,0}^{0}$, which is the collection of the bulk OPE associativity of all $\mathcal{F}_{0,4}$ in the expansion.}
        \label{fig: asso con}
    \end{figure}

\paragraph{Question: Can we express UHP sewing constraints using only bulk correlation functions?}

The UHP sewing constraint consists of four independent conditions originating from the OPE associativity of bulk and boundary operators \cite{Lewellen:1991tb, Runkel:1998he}. We want to ask, can we express the UHP sewing constraints using only bulk correlation functions by applying the linear property?

This should be achievable in principle. We can view the boundary operators $\psi_{i}^{b_{1} b_{2}}$ in the correlation functions as part of the overlaps between two boundaries $\langle b_{2} \vert q^{L_{0} + \Bar{L}_{0} -\frac{c_{Vir}}{12} } \vert b_{1} \rangle$. Then, by expanding two boundary states $\vert b_{1} \rangle$ and $\vert b_{2} \rangle$ in the asymptotic states, we can view $\psi_{i}^{b_{1} b_{2}}$ as part of the OPEs between two infinite sets of bulk operators. The OPE associativity of the boundary operators can be viewed as a certain equivalence in some infinite bulk OPEs. However, we have no equation to support our guess here, so it remains an open question.

\

\section{The single pole structure in the linear coefficients}
\label{sec: exp}

In this section, we attempt to obtain the infinite linear coefficients of expanding Ishibashi states in asymptotic states by direct calculations. The inversion Gram matrices in $I \otimes \Bar{I}$ indicate the existence of singularities at the degenerate limits. We show that the singularities are single poles in the linear coefficients, which is similar to the single pole structure in the chiral conformal block decomposition \cite{Zamolodchikov:1984eqp, Hadasz:2006qb, Hadasz:2008dt, Hadasz:2009db, Hadasz:2012im, Cho:2017oxl}. We calculate the linear coefficients at the lowest levels of Virasoro, $N=1$, $N=2$, and Kac-Moody Ishibashi state expansions to check the single pole structure. 

\

\subsection{The single pole structure at degenerate limit}

Consider a CFT$_2$ with non-heterotic conformal algebra $\mathcal{A} \otimes \mathcal{A}$, with the set of holomorphic conformal generators denoted by $\{ T(z), W^{i}(z), \cdots \}$. An $\mathcal{A}$-module is spanned by the so-called standard basis 
\begin{equation}
    W_{-N^{i}}^{i}  \cdots L_{-N^{0}}  \vert v \rangle , \quad N^{i}:=\{ n_{1}^{i} , n_{2}^{i} , \cdots \} , \quad n_{1}^{i} \ge n_{2}^{i} \ge \cdots\ge 0 , \label{eq: std bas A}
\end{equation}
where $\vert v \rangle$ is an $\mathcal{A}$ primary vector, satisfying 
\begin{equation}
    L_{n} \vert v \rangle = W_{n}^{i} \vert v \rangle =0 ,\quad \forall i , \quad  \forall n \ge 0  .
\end{equation}
An $\mathcal{A}$ primary vector $\vert v \rangle$ is labeled by the eigenvalues (quantum numbers) of the zeroth modes of holomorphic generators with a zero mode
\begin{equation}
    L_{0} \vert v \rangle = h \vert v \rangle, \quad W_{0}^{i} \vert v \rangle = w^{i}  \vert v \rangle .
\end{equation}
Some common examples of quantum numbers include
\begin{itemize}

\item  The holomorphic conformal weight $h$.

    \item The Dynkin labels of the vectors in the highest-weight representations of Kac-Moody algebras.
\end{itemize}
The inner products of the standard basis form the Gram matrix of $\mathcal{A}$-modules. Vectors with different quantum numbers are orthogonal, which can be shown by
\begin{equation}
    \langle v_{2} \vert W_{0}^{i} \vert v_{1} \rangle = w^{i}_{1}\langle v_{2} \vert  v_{1} \rangle = w^{i}_{2}\langle v_{2} \vert  v_{1} \rangle, \quad w_{1}^{i} \ne w_{2}^{i}  \quad  \Rightarrow \quad \langle v_{2} \vert  v_{1} \rangle=0 .
\end{equation}
Hence the infinite-sized Gram matrices are block diagonalized to infinite finite-sized blocks by the quantum numbers.

We view the set of quantum numbers $\{ h, \cdots \}$ as variables. When the determinant of a block of the Gram matrix is zero, then this block can be further block-diagonalized under a type of linear transformation, where one block is a non-zero block and the other is a zero block. This result is based on a \textbf{conjecture} that the norm of the singular vectors $\chi$ at the degenerate limits are proportional to a single zero of the variables
\begin{equation}
    \langle \chi \vert \chi \rangle  \quad \propto \quad  (h-h_{\chi})  \quad \propto \quad (c_{Vir}- c_{\chi}) \quad \propto  \quad \cdots .
\end{equation}
The block diagonalized Gram matrix block takes the form of 
\begin{equation}
    \begin{pmatrix}
    O( \langle \chi \vert \chi \rangle )^{0} &   \cdots  &   O( \langle \chi \vert \chi \rangle )^{0}  &  O( \langle \chi \vert \chi \rangle )^{\frac{1}{2}}  & \cdots  & O( \langle \chi \vert \chi \rangle )^{\frac{1}{2}} \\  \cdots & \cdots & \cdots & \cdots & \cdots & \cdots \\ 
    O( \langle \chi \vert \chi \rangle )^{0} & \cdots &   O( \langle \chi \vert \chi \rangle )^{0} & O( \langle \chi \vert \chi \rangle )^{\frac{1}{2}}  & \cdots  & O( \langle \chi \vert \chi \rangle )^{\frac{1}{2}} \\   O( \langle \chi \vert \chi \rangle )^{\frac{1}{2}} & \cdots & O( \langle \chi \vert \chi \rangle )^{\frac{1}{2}}  &  O( \langle \chi \vert \chi \rangle )^{1}  & \cdots  & O( \langle \chi \vert \chi \rangle )^{1} \\  \cdots  &  \cdots &   \cdots &  \cdots & \cdots &  \cdots  \\  O( \langle \chi \vert \chi \rangle )^{\frac{1}{2}} & \cdots & O( \langle \chi \vert \chi \rangle )^{\frac{1}{2}} &  O( \langle \chi \vert \chi \rangle )^{1} & \cdots & O( \langle \chi \vert \chi \rangle )^{1}
    \end{pmatrix}.
\end{equation}
The above form indicates that all the elements of the inversion of the zero block carry a $(\langle \chi \vert \chi \rangle)^{-1}  $ factor, which is a single pole. When we act the $I \otimes \Bar{I}$ on the Ishibashi states, the holomorphic and antiholomorphic inversion zero block elements contribute totally a double pole, and the inner products between the vectors from degenerate submodules with the Ishibashi state give a first-order zero. Combining the two contributions, we obtain a single pole in the coefficients of vectors from singular submodules.

\ 

We use the Virasoro Ishibashi states $\vert V_{h} (c_{Vir}) \rangle \rangle \in  ( V_{h}  (c_{Vir}) \otimes  U V_{h} (c_{Vir}) ) $ as an example to visualize this single pole structure. The action of $I \otimes \Bar{I}$ on $\vert V_{h} (c_{Vir}) \rangle \rangle$ takes the form of 
\begin{equation*}
    \vert V_{h} (c_{Vir}) \rangle \rangle =  \sum_{\vert M \vert = \vert N \vert =l}^{l \ge 0} \sum_{\vert \Bar{M} \vert = \vert \Bar{N} \vert = \Bar{l} }^{ \Bar{l} \ge 0}  \vert L_{-M} v_{h} \otimes  \Bar{L}_{-\Bar{M}} \Bar{v}_{h} \rangle
\end{equation*}
\begin{equation}
       (G_{c_{Vir},h}^{l})^{MN} (G_{c_{Vir},h}^{\Bar{l}})^{\Bar{M}\Bar{N} }   \langle  L_{-N} v_{h} \otimes  \Bar{L}_{-\Bar{N}} \Bar{v}_{h}  \vert V_{h}  (c_{Vir}) \rangle \rangle . \label{eq: Vir Is exp for}
\end{equation}
The form of the determinant of the level-$l$ Virasoro Gram matrix $(G_{c_{Vir},h}^{l})_{MN}$ was conjectured by Kac and proved by Feigin and Fuchs \cite{Kac:1979fz, Feigin:1981st, Feigin:1982tg}
\begin{equation}
    \det \: (G_{c_{Vir},h}^{l}) \quad \propto \quad  \big[ h-h_{r,s}(c_{Vir})\big]^{p(l-rs)}  ,  \label{eq: lev-l kac det}
\end{equation}
where $p(N)$ is the partition number of integer $N$, $p(N):=0$ when $N<0$. The zeros $h_{r,s}(c_{Vir})$ of the Kac determinant are given by 
\begin{equation}
    h_{r,s} (c_{Vir}) = \frac{c_{Vir}-1}{24}+\frac{1}{4} (r \alpha_{+} + s \alpha_{-} )^2 , \quad \alpha_{\pm}  =  \frac{\sqrt{1-c_{Vir}} \pm  \sqrt{25-c_{Vir}} }{\sqrt{24}} .
\end{equation} 
The form of the Kac determinant indicates that the coefficients of terms from levels $l \ge rs$ are expected to be singular when approaching the degenerate limit $h \to h_{r,s}(c_{Vir})$, where the order of the singularities is yet to be determined. In the sequels, the dependence of $c_{Vir}$ in $h_{r,s}(c_{Vir})$ is omitted.

Next, we normalize the level-$rs$ singular vector $\vert\chi_{rs} \rangle$ as
\begin{equation}
    \vert\chi_{rs} \rangle  = \sum^{M}_{\vert M \vert =rs} \chi_{rs}^{M} \vert L_{-M} v_{r,s} \rangle  , \quad  \chi_{rs}^{ \{  1,  1 ,\cdots ,1 \} } =1 ,
\end{equation}
where $\vert v_{r,s}  \rangle$ is the primary vector of the degenerate Verma module $V_{r,s } (c_{Vir})$. We define vectors $\vert\chi_{rs}^{h} \rangle$ as
\begin{equation}
     \vert \chi_{rs}^{h} \rangle : = \sum_{M}^{\vert M \vert =rs} \chi_{rs}^{M} \vert L_{-M} v_{h} \rangle ,
\end{equation}
where $\vert v_{h} \rangle$ is the Virasoro primary vector of a generic Verma module $V_{h}(c_{Vir})$. The norm of $\vert \chi_{rs}\rangle$ is zero, which is from a first-order zero at the $h \to h_{r,s}$ limit
\begin{equation}
    \lim_{h \to h_{r,s}} \: \langle  \chi_{rs}^{h} \vert  \chi_{rs}^{h} \rangle  = [A_{rs}^{c_{Vir}}]^{-1} (h-h_{r,s}).  \label{eq: nul nor}
\end{equation}
$A_{rs}^{c_{Vir}}$ are non-zero finite numbers, whose forms can be found in \cite{Zamolodchikov:2003yb, Cho:2017oxl}. An immediate conclusion following (\ref{eq: nul nor}) is that the norms of $\vert L_{-M} \chi_{rs} \rangle$ process the same first-order zero 
\begin{equation}
      \lim_{h \to h_{r,s}} \: \langle  L_{-M} \chi_{rs}^{h} \vert  L_{-M} \chi_{rs}^{h} \rangle  = [A_{rs}^{c_{Vir}}]^{-1} (h-h_{r,s}) \times \langle  L_{-M} v_{h_{r,s}+rs}  \vert L_{-M} v_{h_{r,s}+rs}  \rangle.
\end{equation}

We define the level-$l$ singular basis for the Virasoro Verma module by setting the last $p(l-rs)$ basis vectors as level-$(l-rs)$ descendants in the singular submodule $\vert L_{-N} \chi_{rs} \rangle$, $\vert N \vert = l-rs$. Linear transforming from the standard basis to this basis block diagonalizes the level-$l$ Gram matrix to a $ [p(l)-p(l-rs)]  \times [p(l)-p(l-rs)]  $ sized non-zero block and a $p(l-rs) \times p(l-rs)$ sized zero block
\begin{equation}
    \lim_{h \to h_{r,s}} (S_{c_{Vir},h}^{l} )_{ab} = \begin{pmatrix}
    O(h-h_{r,s})^{0} &   \cdots  &    O(h-h_{r,s})^{0}  &  O(h-h_{r,s})^{\frac{1}{2}}  & \cdots  & O(h-h_{r,s})^{\frac{1}{2}} \\  \cdots & \cdots & \cdots & \cdots & \cdots & \cdots \\ 
     O(h-h_{r,s})^{0} & \cdots &    O(h-h_{r,s})^{0} & O(h-h_{r,s})^{\frac{1}{2}}  & \cdots  & O(h-h_{r,s})^{\frac{1}{2}} \\   O(h-h_{r,s})^{\frac{1}{2}} & \cdots &  O(h-h_{r,s})^{\frac{1}{2}}  &  O(h-h_{r,s})^{1}  & \cdots  & O(h-h_{r,s})^{1} \\  \cdots  &  \cdots &   \cdots &  \cdots & \cdots &  \cdots  \\  O(h-h_{r,s})^{\frac{1}{2}} & \cdots & O(h-h_{r,s})^{\frac{1}{2}} &  O(h-h_{r,s})^{1} & \cdots & O(h-h_{r,s})^{1}
    \end{pmatrix}. \label{eq: Gr ma nul bas}
\end{equation}
The determinant of $(S_{c_{Vir},h_{r,s}}^{l} )_{ab}$ is at the order $O\big[(h-h_{r,s})^{p(l-rs)}\big]$, which is the same as the determinant zero block. This indicates that the inversion of the non-zero block is well-defined, and the zero block is singular.

We rewrite the identity operator $I \otimes \Bar{I}$ under the singular basis. Since all the elements in the zero block are at the order of $O([h-h_{r,s}])$, all elements in the inversion of the zero block are at the order of $O([h-h_{r,s}]^{-1})$. Hence, the two inversion Gram matrices contribute a total $(h-h_{r,s})^{-2} $ singularity for all terms from the singular submodule. The inner product between vectors from the singular submodule and the Ishibashi states will contribute a first-order zero $(h-h_{r,s})$ from the norm of vectors from the singular submodule. Combining all contributions, we obtain the order of singular terms at $(h-h_{r,s})^{-1}$. 

\

\subsection{The Ishibashi state expansion at the lowest levels} \label{subsec: Is exp pra}

We demonstrate some simple Ishibashi state expansion at the lowest levels to check the single pole structure. We will expand generic Virasoro, $N=1$, $N=2$, and $\widehat{su(2)}_{k}$ Kac-Moody Ishibashi states at the lowest levels. We also show that in some special cases, the complete expansion up to level-$\infty$ can be obtained easily, such as the free field CFT$_2$ and the $c_{Vir} \to \infty$ semi-classical limit. 

\ 

\subsubsection{Virasoro Ishibashi states}

The commutation relations if the Virasoro algebra are given by \cite{Belavin:1984vu}
\begin{equation}
    [L_{m},L_{n}]=(m-n)L_{m+n}+ \frac{ c_{Vir} }{12}(m^3-m)\delta_{m+n,0} ,
\end{equation}
where the Virasoro generators $L_{n}$ are the Laurent modes of the energy-stress tensor
\begin{equation}
    T(z) = \sum_{n\in \mathbb{Z} } \frac{ L_{n} }{z^{n+2}}  , \quad L_{n} = \oint \frac{dz}{2\pi i} z^{n+1}  T(z) .
\end{equation}

Consider the expansion of a generic Virasoro Ishibashi state $\vert V_{h} \rangle \rangle \in V_{h} \otimes U V_{h} $. At level-$0$ and level-$1$, the expansion is simple 
\begin{equation}
    \vert v_{h} \otimes U \Bar{v}_{h} \rangle + \frac{1}{2h} \vert  L_{-1} v_{h} \otimes U  \Bar{L}_{-1}\Bar{v}_{h} \rangle .
\end{equation}
The coefficient for the $\vert  L_{-1} v_{h} \otimes U  \Bar{L}_{-1} \Bar{v}_{h} \rangle$ term has a single pole at the $h \to h_{1,1}=0$ limit, which is from $(\langle \chi_{1}^{h} \vert \chi_{1}^{h} \rangle  )^{-1}$. The level-$2$ expansion is less simple. Nevertheless, the result can be neatly summarized as
\begin{equation}
    \frac{9}{64(h+\frac{1}{2}) (h-h_{2,1}) (h-h_{1,2})} \vert \chi_{2}^{h} \otimes U \Bar{\chi}_{2}^{h} \rangle + \frac{1}{8(h-h_{1,1})(h+\frac{1}{2})}  \vert L_{-1} \chi_{1}^{h}  \otimes  U  \Bar{L}_{-1}  \Bar{\chi}_{1}^{h}  \rangle,
\end{equation}
where the vector $\vert \chi_{2}^{h} \rangle$ is 
\begin{equation}
    \vert \chi_{2}^{h} \rangle = \Big[  - \frac{2(2h+1)}{3} L_{-2}  + L_{-1}^{2} \Big] \vert v_{h} \rangle.
\end{equation}
The single poles at the $h \to h_{2,1}$ and $h_{1,2}$ limits are from $(\langle \chi_{2}^{h} \vert \chi_{2}^{h} \rangle)^{-1}$. The single pole at the $h \to h_{1,1}$ limit is from $ ( \langle L_{-1} \chi_{1}^{h}  \vert L_{-1} \chi_{1}^{h} \rangle )^{-1}$.

At level-$3$, we do not write down the detailed form of expansion. However, it is easy to check the existence of single poles $(h-h_{3,1})^{-1}$ and $(h-h_{1,3})^{-1}$. We show that at the degenerate limits $h \to h_{3,1}$ and $h_{1,3}$, one of the three orthonormal vectors is proportional to $\vert \chi_{3}^{h} \rangle$ 
\begin{equation*}
    \lim_{h \to h_{3,1} \: \text{or} \: h_{1,3}  } \Big[ \vert L_{-1}^{3} v_{h}  \rangle +   \frac{2(5h+1)(2h-1)}{3+c_{Vir}-7h}  \vert  L_{-2} L_{-1} v_{h} \rangle 
\end{equation*}
\begin{equation*}
     + \frac{c_{Vir} (1 + h) (1 + 2 h) + 2 (-1 + h) (-1 + h + 8 h^2)}{7h-c_{Vir}-3}  \vert L_{-3} v_{h}  \rangle  \Big] = \vert \chi_{3}^{h} \rangle,
\end{equation*}
where $\vert \chi_{3}^{h} \rangle$ is
\begin{equation}
    \vert \chi_{3}^{h} \rangle = \vert L_{-1}^{3}  v_{h} \rangle -2(h+1) \vert L_{-2}L_{-1} v_{h} \rangle + h(h+1) \vert L_{-3} v_{h} \rangle .
\end{equation}

We also consider expressing the linear coefficients using the Virasoro central charge $c_{Vir}$ as the variable. The zeros of the Kac determinant expressed in Virasoro central charges are
\begin{equation}
    c_{r,s}= 1 +6 \big[ b_{r,s}(h) + b_{r,s}(h)^{-1}   \big]^{2}
\end{equation}
where $b= i \alpha_{+}$ and
\begin{equation}
    b_{r,s}(h)^{2} = \frac{rs-1+2h + \sqrt{(r-s)^{2} +4(rs-1)h +4h^{2} } }{1-r^{2}} , \quad r\ge 2 , \quad s \ge 1 .
\end{equation}
The level-$2$ expansion can be expressed as 
\begin{equation}
    \frac{2}{c_{Vir}-c_{2,1}(h) } \vert \chi_{2}^{h} \otimes U \Bar{\chi}_{2}^{h} \rangle + \frac{1}{4h(2h+1)}  \vert L_{-1} \chi_{1}^{h}  \otimes  U  \Bar{L}_{-1}  \Bar{\chi}_{1}^{h}  \rangle ,
\end{equation}
where $c_{2,1}(h)$ is
\begin{equation}
    c_{2,1}(h)=c_{1,2}(h) = 9-8h -\frac{9}{2h+1} .
\end{equation}

\paragraph{Virasoro Ishibashi state expansion at the $c_{Vir} \to \infty$ limit:}

We end the practice of Virasoro Ishibashi state expansions by considering the semi-classical limit $c_{Vir} \to \infty$, with $h$ fixed to be finite. At this limit, the non-zero elements in the inversion Gram matrices are $(G_{c_{Vir} ,h }^{l})^{L_{-1}^{l} , L_{-1}^{l} }$, $l \in \mathbb{N}$, which indicates that we only need to calculate the norms of $\vert L_{-1}^{l} v_{h}\rangle$ to determine the full Virasoro Ishibashi state expansion up to level-$\infty$ at the $c_{Vir} \to \infty$ limit
\begin{equation*}
    \langle L_{-1}^{l} v_{h} \vert L_{-1}^{l} v_{h} \rangle = l! (2h)_{l},
\end{equation*}
\begin{equation}
    \lim_{c_{Vir} \to \infty }  \vert V_{h} \rangle \rangle =  \sum_{l=0}^{\infty} \frac{1}{l! (2h)_{l}} \vert L_{-1}^{l} v_{h} \otimes U \Bar{L}_{-1}^{l} \Bar{v}_{h} \rangle ,
\end{equation}
where $a_{l}:= a(a+1)\cdots(a+l-1)$ are the Pochhammer symbols. The single poles in coefficients only exist at the $h \to h_{1,1}$ limit, which are from $( \langle  L_{-1}^{l} v_{1,1} \vert  L_{-1}^{l} v_{1,1} \rangle )^{-1}$, $l \in \mathbb{Z}^{+} $. 

\ 

\subsubsection{$N=1$ Ishibashi states}

The $N=1$ superconformal algebra is generated by two holomorphic generators: the energy-stress tensor $T(z)$ and the supercurrent $T_{F}(z)$. The (anti-)commutation relations of the $N=1$ algebras are given by
\begin{equation*}
    [L_{m},L_{n}]=(m-n)L_{m+n}+ \frac{\Hat{c}}{8}(m^3-m)\delta_{m+n,0} , \quad \Hat{c}= \frac{2}{3}c_{Vir} ,
\end{equation*}
\begin{equation}
    [L_{m},G_{r}]= \Big( \frac{m}{2}-r \Big) G_{m+r} , \quad  \{ G_{r} ,G_{s} \} =2 L_{r+s} + \frac{\Hat{c}}{2}\Big( r^{2}-\frac{1}{4} \Big) \delta_{r+s,0} .
\end{equation}
where
\begin{equation}
    L_{n} =\oint \frac{dz}{2\pi i} z^{n+1} T(z)  , \quad G_{r} = 2 \oint \frac{dz}{2\pi i} z^{r+\frac{1}{2}}  T_{F}(z) .
\end{equation}
The choice of $r\in \mathbb{Z}+\frac{1}{2}$ is called the NS sector and the choice of $r \in \mathbb{Z}$ is called the R sector \cite{Friedan:1984rv, Bershadsky:1985dq}. The zeros in both the NS and R sector $N=1$ Kac determinants $h_{r,s}^{N=1} (\Hat{c}) $ are given by
\begin{equation}
    h_{r,s}^{N=1} (\Hat{c}) = \frac{\Hat{c}-1}{16} + \frac{1-(-1)^{r-s}}{32} + \frac{1}{8} \Big( r\alpha_{+}^{N=1} + s \alpha_{-}^{N=1} \Big)^{2}  ,  \quad  \alpha_{\pm}^{N=1} = \frac{\sqrt{ 1-\Hat{c} }  \pm  \sqrt{9-\Hat{c}} }{\sqrt{8}} ,
\end{equation}
where $\vert r-s \vert \in 2\mathbb{Z}$ belong to the NS sector, while the zeros with $\vert r-s \vert \in 2\mathbb{Z}+1$ belong to the R sector. When $h \to h_{r,s}^{N=1} (\Hat{c}) $, the $N=1$ module process a level-$rs/2$ singular vector. $N=1$ Ishibashi states $\vert V_{h}^{N=1} \rangle \rangle_{\pm}$ allow two types of gluing automorphisms $ \Omega= \pm 1 $, corresponding to the two types of gluing conditions
\begin{equation}
   \big( L_{n} - \Bar{L}_{-n}  \big) \vert V_{h}^{N=1} \rangle \rangle_{\pm} = \big( G_{r} \pm i \Bar{G}_{-r}  \big) \vert V_{h}^{N=1} \rangle \rangle_{\pm} =0.
\end{equation}

\paragraph{NS sector expansion:}

We expand a generic NS $N=1$ Ishibashi state $\vert V_{h}^{\text{NS} ;N=1} \rangle \rangle_{\pm}$ up to level-$\frac{3}{2}$. The expansion up to level-$1$ is simple
\begin{equation}
     \vert v_{h} \otimes U \Bar{v}_{h}  \rangle + \frac{1}{2(h-h_{1,1}^{N=1})} \vert G_{-\frac{1}{2}} v_{h} \otimes (\pm U)  \Bar{G}_{-\frac{1}{2}}  \Bar{v}_{h} \rangle + \frac{1}{2(h-h_{1,1}^{N=1})} \vert L_{-1} v_{h} \otimes U \Bar{L}_{-1} \Bar{v}_{h} \rangle .
\end{equation}
The single poles at the $h \to h_{1,1}^{N=1}$ limit are from $(\langle \chi_{\frac{1}{2}}^{h; N=1  } \vert \chi_{\frac{1}{2}}^{h;N=1   } \rangle)^{-1}$ and \\ $(\langle G_{-\frac{1}{2}} \chi_{\frac{1}{2}}^{h;N=1  } \vert  G_{-\frac{1}{2}}\chi_{\frac{1}{2}}^{h;N=1  } \rangle)^{-1}$, where $\vert \chi_{\frac{1}{2}}^{h;N=1  } \rangle =  G_{-\frac{1}{2}} \vert v_{h} \rangle$. The less simple level-$\frac{3}{2}$ expansion can be summarized as 
\begin{equation*}
     \frac{(h+\frac{\Hat{c}}{2})}{ 4(h-h_{1,1}^{N=1}) (h-h_{3,1}^{N=1}) (h-h_{1,3}^{N=1})} \vert \chi_{\frac{3}{2}}^{h; N=1  } \otimes (\pm U) \Bar{\chi}_{\frac{3}{2}}^{h;N=1  } \rangle 
\end{equation*}
\begin{equation}
    +\frac{1}{(2h+\Hat{c})} \vert G_{-\frac{3}{2}} v_{h}  \otimes  (\pm  U)  \Bar{G}_{-\frac{3}{2}} \Bar{v}_{h}  \rangle.
\end{equation}
The single poles at $h \to h_{1,1}^{N=1}$, $h_{3,1}^{N=1}$ and $h_{1,3}^{N=1}$ limits are from $(\langle \chi_{\frac{3}{2}}^{h; N=1  } \vert \chi_{\frac{3}{2}}^{h; N=1  } \rangle)^{-1}$, where 
\begin{equation}
    \vert \chi_{\frac{3}{2}}^{h; N=1  } \rangle = \Big[ \frac{ 4h }{ (\Hat{c}+ 2h) } G_{-\frac{3}{2}} -  G_{-\frac{1}{2}} L_{-1}  \Big] \vert v_{h} \rangle.
\end{equation}

We also express the level-$\frac{3}{2}$ expansion coefficients using $\Hat{c}$ as the parameter 
\begin{equation}
    \frac{(h+\frac{\Hat{c}}{2})}{ \big[\Hat{c}-\Hat{c}_{3,1}(h) \big] h(2h+1) } \vert \chi_{\frac{3}{2}}^{h; N=1 } \otimes (\pm U) \Bar{\chi}_{\frac{3}{2}}^{h;N=1  } \rangle +\frac{1}{(2h+\Hat{c})} \vert G_{-\frac{3}{2}} v_{h}  \otimes  (\pm  U)  \Bar{G}_{-\frac{3}{2}} \Bar{v}_{h}  \rangle,
\end{equation}
where $\Hat{c}_{3,1}=\Hat{c}_{1,3}=\frac{2h(3-2h)}{2h+1}$.

\paragraph{The $ \Hat{c} \to \infty$ limit of NS $N=1$ Ishibashi state expansions: }

At the $ \Hat{c} \to \infty$ limit, the non-zero contributions to NS $N=1$ Ishibashi state expansion are from $\vert L_{-1}^{l} v_{h} \rangle$ and $\vert G_{-\frac{1}{2}} L_{-1}^{l} v_{h} \rangle $. We only need to compute the norms of those vectors to fully determine the NS $N=1$ Ishibashi state expansion up to level-$\infty$ at the $\Hat{c} \to \infty$ limit
\begin{equation}
    \lim_{\Hat{c} \to \infty}\vert V_{h}^{ \text{NS} ;N=1} \rangle \rangle_{\pm} = \sum_{l=0}^{\infty} \Big[ \frac{\vert L_{-1}^{l} v_{h} \otimes U   \Bar{L}_{-1}^{l} \Bar{v}_{h}\rangle}{l!(2h)_{l}} + \frac{\vert G_{-\frac{1}{2}} L_{-1}^{l} v_{h} \otimes  ( \pm U ) \Bar{G}_{-\frac{1}{2}} \bar{L}_{-1}^{l} \Bar{v}_{h}\rangle}{l!(2h)_{l+1}} \Big] .
\end{equation}
The poles at the $h \to h_{1,1}^{N=1}=0$ limit are from $(\langle L_{-1}^{l+1} v_{h} \vert L_{-1}^{l+1} v_{h} \rangle)^{-1}$ and $(\langle G_{-\frac{1}{2}} L_{-1}^{l} v_{h} \vert G_{-\frac{1}{2}} L_{-1}^{l} v_{h} \rangle)^{-1}$, $l\in \mathbb{N}$.

\paragraph{R sector expansions:}

In the R sector, the Hilbert space is separated into the fermion number $\pm$ subsectors. Fermion numbers $F$ are determined by the total number of supercurrent modes that act on the primary bosonic vector $\vert v_{h}^{+}\rangle$.

The expansion of the R sector $N=1$ Ishibashi state at level-$0$ is 
\begin{equation}
     \vert v_{h}^{+} \otimes U \Bar{v}_{h}^{+} \rangle + \frac{1}{(h-\frac{\Hat{c}}{16})}  \vert v_{h}^{-} \otimes  ( \pm U ) \Bar{v}_{h}^{-} \rangle,
\end{equation}
where $\vert v_{h}^{-} \rangle:= G_{0}  \vert v_{h}^{+}\rangle$ is a singular vector when $h= \frac{\Hat{c}}{16}$. The fermion number $+$ expansion at level-$1$ is
\begin{equation*}
    \frac{(h+ \frac{3\Hat{c}}{16})}{2(h-h_{2,1}^{N=1}) (h-h_{1,2}^{N=1})  }  \vert \chi_{1}^{h ; N=1 , +} \otimes U \Bar{\chi}_{1}^{h ; N=1 , +} \rangle
\end{equation*}
\begin{equation}
       + \frac{1 }{2(h+\frac{3\Hat{c}}{16})(h-\frac{\Hat{c}}{16})}  \vert  G_{-1} v_{h}^{-}  \otimes   U  \Bar{G}_{-1} \Bar{v}_{h}^{-} \rangle ,
\end{equation}
where $\vert \chi_{1}^{h ;  N=1 ,+} \rangle$ is 
\begin{equation}
    \vert \chi_{1}^{h ; N=1 , +} \rangle = - \frac{3}{4(h+\frac{3\hat{c}}{16})}  G_{-1} \vert v^{-}_{h} \rangle +  L_{-1} \vert v^{+}_{h} \rangle .
\end{equation}
The poles at the $h \to h_{2,1}^{N=1}$ and $h_{1,2}^{N=1}$ limits are from $( \langle\chi_{1}^{h ; N=1 , +}  \vert \chi_{1}^{h ;N=1 , +} \rangle)^{-1}$. The fermion number $-$ expansion at level-$1$ is
\begin{equation*}
     \frac{(h+ \frac{3\Hat{c}}{16})}{2(h-h_{2,1}^{N=1}) (h-h_{1,2}^{N=1})(h-\frac{\Hat{c}}{16})} \vert \chi_{1}^{h; N=1 , -} \otimes (\pm U) \Bar{\chi}_{1}^{h;N=1 , -} \rangle  
\end{equation*}
\begin{equation}
   + \frac{1}{2(h+\frac{3\Hat{c}}{16})} \vert G_{-1} v^{+}_{h} \otimes (\pm U) \Bar{G}_{-1} \Bar{v}^{+}_{h} \rangle,
\end{equation}
where $\vert \chi_{1}^{h;  N=1 , -} \rangle$ is 
\begin{equation}
    \vert \chi_{1}^{h; N=1 , -} \rangle  = \vert L_{-1} v^{-}_{h} \rangle - \frac{3 (h- \frac{ \Hat{c} }{16} ) }{4(h+ \frac{3 \Hat{c} }{16} )} \vert G_{-1} v^{+}_{h} \rangle .
\end{equation}
The poles at the $h_{2,1}^{N=1}$, and $h_{1,2}^{N=1}$ limits are from $ ( \langle\chi_{1}^{h; N=1 , -}  \vert \chi_{1}^{h; N=1 , -} \rangle )^{-1}$.

\paragraph{The $\Hat{c} \to \infty$ limit of the $N=1$ R Ishibashi state expansions:}

At the $\Hat{c} \to \infty$ limit, the non-zero contributions in the R sector Ishibashi state expansions are from $\vert L_{-1}^{l} v_{h}^{+} \rangle$. The $N=1$ R sector Ishibashi state expansion at $\Hat{c} \to \infty$ limit is
\begin{equation}
    \lim_{ \Hat{c} \to \infty } \vert V_{h}^{N=1, \text{R}} \rangle \rangle_{\pm} = \sum_{l=0}^{\infty} \: \frac{ 1}{l!(2h)_{l}}   \vert L_{-1}^{l} v_{h}^{+} \otimes  U  \Bar{L}_{-1}^{l} \Bar{v}_{h}^{+} \rangle  .
\end{equation}

\

\subsubsection{$N=2$ NS Ishibashi states}

The $N=2$ superconformal algebra is generated by four conformal generators, namely the energy-stress tensor $T(z)$, a $U(1)$ current $J(z)$, and two types of supercurrents $G^{\pm}(z)$ with $U(1)$ charges $\pm 1$ respectively \cite{DiVecchia:1985ief, DiVecchia:1986cdz, DiVecchia:1986fwg}. The (anti-)commutation relations in the $N=2$ algebra are
\begin{equation*}
    [L_{m},L_{n}]= \frac{c_{vir}}{12} (m^{3}-m) \delta_{m+n,0} + (m-n)L_{m+n}, \quad [L_{m}, G_{n\pm a }^{\pm} ] = \big(  \frac{m}{2}-n \mp a \big) G_{m+n\pm a}^{\pm},
\end{equation*}
\begin{equation*}
    [L_{m}, J_{n}]= -n J_{m+n}  , \quad [J_{m}, J_{n}]= \frac{c_{Vir}}{3}m  \delta_{m+n,0}, \quad [J_{m}, G^{\pm}_{n\pm a}] = \pm G_{m+n \pm a}^{\pm} , 
\end{equation*}
\begin{equation*}
    \{ G_{m+a}^{+} , G_{n-a}^{-}  \} = 2L_{m+n} + (m-n+2a)J_{m+n} + \frac{c_{Vir}}{3} \Big[ (m+a)^{2}-\frac{1}{4} \Big] \delta_{m+n} , \quad 
\end{equation*}
\begin{equation}
   \{ G_{m+a}^{+} ,  G_{n+a}^{+}  \} = \{ G_{m-a}^{-} , G_{n-a}^{-}  \} = 0. 
\end{equation}
The choice of $a$ will determine the sector of the algebra, where $a=0$ and $a=\frac{1}{2}$ corresponding to the R and NS sectors respectively.

We demonstrate the expansion of $N=2$ NS sector twisted Ishibashi states $\vert V_{h,q} \rangle \rangle_{\omega}$ up to level-$1$. Two types of gluing conditions are allowed, which are the so-called A-type and B-type gluing conditions \cite{Ooguri:1996ck}. For simplicity, we don't write down the detailed action of $V_{\omega}$ on the conformal generator.

The expansion of $\vert V_{h,q} \rangle \rangle_{\omega}$ up to level-$\frac{1}{2}$ is
\begin{equation}
    \vert v_{h,q} \otimes V_{\omega} U v_{h,q} \rangle +  \frac{ \vert G_{-\frac{1}{2}}^{+}   v_{h,q} \otimes V_{\omega} U \Bar{G}_{-\frac{1}{2}}^{+} \Bar{v}_{h,q} \rangle}{(2h-q)}  +  \frac{\vert G_{-\frac{1}{2}}^{-}   v_{h,q} \otimes V_{\omega} U \Bar{G}_{-\frac{1}{2}}^{-} \Bar{v}_{h,q} \rangle}{(2h+q)} . 
\end{equation}
The poles at $h \to  \pm \frac{q}{2}$ are from $ ( \langle   G_{-\frac{1}{2}}^{\mp} v_{h,q} \vert G_{-\frac{1}{2}}^{\mp} v_{h,q} \rangle)^{-1} $. The level-$1$ expansion is 
\begin{equation*}
     \Big\{  \frac{2(c_{Vir}-3)h +2c_{Vir}-(3+c_{Vir})q}{(2h-q) [2c_{Vir}h+c_{Vir}-3(2h+q^2) ]}\Big\} \vert \chi_{1}^{N=2; \text{NS} ,1} \otimes V_{\omega} U  \Bar{\chi}_{1}^{N=2; \text{NS} ,1}  \rangle
\end{equation*}
\begin{equation*}
    + \Big[   \frac{3(2+2h-q)}{[2(c_{Vir}-3)h +2c_{Vir}-(3+c_{Vir})q]} \Big]  \vert \chi_{1}^{N=2; \text{NS} ,2} \otimes V_{\omega} U  \Bar{\chi}_{1}^{N=2; \text{NS} ,2}  \rangle
\end{equation*}
\begin{equation}
    +  \frac{1}{(2h+q)(2+2h-q)}  \vert G_{-\frac{1}{2}}^{+} G_{-\frac{1}{2}}^{-} v_{h,q} \otimes V_{\omega} U \Bar{G}_{-\frac{1}{2}}^{+} \Bar{G}_{-\frac{1}{2}}^{-} \Bar{v}_{h,q} \rangle , 
\end{equation}
where
\begin{equation*}
   \vert \chi_{1}^{N=2; \text{NS} ,1} \rangle = \vert L_{-1} v_{h,q} \rangle - \frac{3(2h-q)(q-1)  }{2(c_{Vir}-3)+2c_{Vir}-q(c_{Vir}+3)}  \vert J_{-1} v_{h,q} \rangle 
\end{equation*}
\begin{equation}
    - \frac{(c_{Vir}-3q) }{2(c_{Vir}-3)+2c_{Vir}-q(c_{Vir}+3)} \vert G_{-\frac{1}{2}}^{+} G_{-\frac{1}{2}}^{-} v_{h,q} \rangle ,
\end{equation}
and
\begin{equation}
   \vert \chi_{1}^{N=2; \text{NS} ,2} \rangle =  \Big[ J_{-1} - \frac{1}{(2+2h-q)}G_{-\frac{1}{2}}^{+} G_{-\frac{1}{2}}^{-} \Big] \vert v_{h,q} \rangle . 
\end{equation}
The poles at the $h \to \frac{q}{2}$ and $h \to \frac{3q^2-c_{Vir}}{2(c_{Vir}-3)}$ limits are from $( \langle  \chi_{1}^{N=2; \text{NS} ,1} \vert \chi_{1}^{N=2; \text{NS} ,1} \rangle)^{-1}$. The pole at the $h \to \frac{c_{Vir}q+3q-2c_{Vir}}{2(c_{Vir}-3)}$ limit is from $( \langle \chi_{1}^{N=2;\text{NS} ,2}  \vert \chi_{1}^{N=2; \text{NS} ,2} \rangle)^{-1}$. The poles at the $h \to -\frac{q}{2}$ and $h \to \frac{q-2}{2}$ limits are from $(\langle  G_{-\frac{1}{2}}^{+} G_{-\frac{1}{2}}^{-} v_{h,q} \vert G_{-\frac{1}{2}}^{+} G_{-\frac{1}{2}}^{-} v_{h,q} \rangle )^{-1}$.

\paragraph{The $c_{Vir} \to \infty$ limit of the $N=2$ NS Ishibashi state expansion:}

The non-zero elements that contribute to $N=2$ NS Ishibashi state at the $c_{Vir} \to \infty $ limit are $ L_{-1}^{l} \vert v_{h,q} \rangle$, $ G_{-\frac{1}{2}}^{+} L_{-1}^{l} \vert v_{h,q} \rangle$, $G_{-\frac{1}{2}}^{-} L_{-1}^{l} \vert v_{h,q} \rangle$, and $G_{-\frac{1}{2}}^{+} G_{-\frac{1}{2}}^{-} L_{-1}^{l}\vert v_{h,q} \rangle$. The basis vectors at level-$(l+\frac{1}{2})$ are orthogonal, but the basis vectors at level-$(l+1)$ are not. We need to compute the explicit level-$(l+1)$ Gram matrices to obtain the expansion. The result of the expansion is
\begin{equation*}
    \lim_{c_{Vir} \to \infty}  \vert V_{h,q} \rangle \rangle =  \sum_{l=0}^{\infty} \frac{\vert G_{-\frac{1}{2}}^{+} L_{-1}^{l} v_{ h, q} \otimes  V_{\omega} U  \Bar{G}_{-\frac{1}{2}}^{+} \Bar{L}_{-1}^{l} \Bar{v}_{ h, q} \rangle }{l!(2h-q)(2h)_{l} } + \frac{\vert G_{-\frac{1}{2}}^{-} L_{-1}^{l} v_{ h, q} \otimes  V_{\omega} U  \Bar{G}_{-\frac{1}{2}}^{-} \Bar{L}_{-1}^{l} \Bar{v}_{ h, q} \rangle }{l!(2h+q)(2h)_{l} } +
\end{equation*}
\begin{equation}
     \frac{\vert L_{-1}^{l} v_{h,q} \otimes  V_{\omega} U   \Bar{L}_{-1}^{l} \Bar{v}_{h,q}  \rangle }{ l! (2h)_{l}}  + \frac{(2h+q) \vert \chi_{l+1}^{N=2, \text{NS}, c_{Vir}\to \infty} \otimes V_{\omega} U \Bar{\chi}_{l+1}^{N=2, \text{NS}, c_{Vir}\to \infty} \rangle }{2h[4h^2+2h(2+l-q)+ lq ] l! (2h)_{l} - (l+1)! (2h)_{l+1}} ,
\end{equation}
where $\vert\chi_{l+1}^{N=2, \text{NS}, c_{Vir}\to \infty} \rangle$ are 
\begin{equation}
   \vert \chi_{l+1}^{N=2, \text{NS}, c_{Vir}\to \infty} \rangle := \Big[ L_{-1}^{l+1}  - \frac{2h}{2h+q}  G_{-\frac{1}{2}}^{+} G_{-\frac{1}{2}}^{-} L_{-1}^{l} \Big]  \vert v_{h,q} \rangle. 
\end{equation}
The poles exist at $h\to 0$, $h \to \pm q/2$, and $h \to h_{l}$ limits, where $h_{l}$ are the solutions of equations
\begin{equation}
    2h[4h^2+2h(2+l-q)+ lq ]   =(l+1) (2h+l), \quad l \in \mathbb{Z}_{+} .
\end{equation}
If we further set $q \to \infty$, the expansion can be further simplified
\begin{equation}
    \lim_{c_{Vir},q \to \infty} \vert V_{h,q} \rangle \rangle =  \sum_{l=0}^{\infty}\frac{\vert L_{-1}^{l} v_{h,q} \otimes  V_{\omega} U   \Bar{L}_{-1}^{l} \Bar{v}_{h,q}  \rangle }{l! (2h)_{l}}  + \frac{\vert L_{-1}^{l} v_{h,q} \otimes  V_{\omega} U   \Bar{L}_{-1}^{l} \Bar{v}_{h,q}  \rangle}{2h(l-2h)l! (2h)_{l}} .
\end{equation}

\

\subsubsection{$\widehat{su(2)}_{k}$ Kac-Moody Ishibashi states}

Consider a finite-dimensional Lie algebra $g$ with commutation relation
\begin{equation}
    [J^{a} , J^{b}] = \sum_{c} i f^{ab}_{c} J^{c},
\end{equation}
where $J^{i}$ are $g$ generators, and $f^{ab}_{c}$ are the structure constants. The corresponding Kac-Moody algebra $\Hat{g}_{k}$ is 
\begin{equation}
    [J^{a}_{m} , J^{b}_{n}  ]  =  k m \delta_{ab} \delta_{m+n,0}  + i f^{ab}_{c}  J^{c}_{m+n}   ,  \quad J^{a}_{m}:= J^{a} \otimes z^{m} , \quad z \in \mathbb{C}, 
\end{equation}
where $k$ is the Kac-Moody level. The descendant vectors Kac-Moody modules are 
\begin{equation}
    L_{-N} J_{-N_{a}}^{a} \cdots \vert v \rangle.
\end{equation}
Because the Virasoro generators $L_{m}$ can be obtained from the Sugawara constructions of the current modes $J^{a}_{m}$. Hence, Kac-Moody modules are spanned by the current descendants $  J^{a}_{-N_{a}} \cdots \vert v \rangle$, which are used as our basis.

\paragraph{Abelian Kac-Moody Ishibashi state expansion:}

Consider a Kac-Moody algebra $\Hat{h}$ related to a finite Abelian Lie algebra $h \cong u(1)^{ \dim{h} }$, with generators $H^{a}$, $a=1,\cdots, \dim \: h$. Abelian current algebras have simple commutation relations 
\begin{equation}
     [H_{m}^{a} , H_{n}^{b}]=km \delta_{ab} \delta_{m+n,0},
\end{equation}
where the level $k$ can always be absorbed by a rescale of the generators $H^{a} \to  \Tilde{H}^{a} = H^{a} / \sqrt{k} $. From the commutation relation, we can show that the current descendant basis  $ \prod_{a=1}^{\dim\: h} \Tilde{H}_{-N_{a}}^{a} \vert v \rangle $ is orthogonal. Hence, we only need to calculate the norms of all current descendants to obtain the full Ishibashi state expansion up to level-$\infty$
\begin{equation}
    \Big\langle \prod_{a=1}^{\dim h } \Tilde{H}_{-N^{a}}^{a}  v \Big\vert  \prod_{a=1}^{\dim h } \Tilde{H}_{-N^{a}}^{a}  v \Big\rangle  =\prod_{a=1}^{\dim h }  \prod_{i} m_{i}^{a}! (n_{i}^{a})^{m_{i}^{a}} ,\quad N^{a} = \{  (n_{i}^{a} )^{m_{i}^{a}} , \cdots   \} . \label{eq: nor Abe cur}
\end{equation}
From (\ref{eq: nor Abe cur}), the Abelian Kac-Moody Ishibashi state expansions up to level-$\infty$ can be summarized as
\begin{equation}
    \vert V \rangle \rangle_{\omega}  =  \exp{ \Big\{ \sum_{i=1}^{\dim  h} \sum_{n=1}^{\infty}  \frac{1}{n} \Tilde{H}_{-n}^{a}  V_{\omega} U  \Bar{\Tilde{H}}_{-n}^{a}  \Big\}  } \vert v \otimes \Bar{v}\rangle.
\end{equation}

When $\dim \: h =1$, we recover the celebrated free boson Ishibashi states \cite{Callan:1987px}. For uncompactified free boson CFT, the Ishibashi states for the momentum $p$ $U(1)$ Fock space are 
\begin{equation}
    \vert \alpha \rangle \rangle_{D, N} = \exp \Big\{ \sum_{n=1}^{\infty} \mp \frac{1}{n} \alpha_{-n} \Bar{\alpha}_{-n}  \Big\} \vert \alpha ,\bar{\alpha} \rangle.  \label{eq: fre bos Ishi}
\end{equation}
For Dirichlet Ishibashi states, the gluing condition requires $\alpha= \Bar{\alpha}$. For Neumann Ishibashi states, the gluing condition requires $\alpha + \Bar{\alpha}=0$.

Consider a compactified free boson CFT, whose Fock spaces are labeled by momentum $p$ and winding number $w$. The Dirichlet boundary condition requires zero winding number $(J_{0}-\Bar{J}_{0}) \vert (p,w) \rangle \rangle_{D} = \Hat{w} \vert (p,w) \rangle \rangle_{D} =0$. Similarly, the Neumann boundary condition requires zero momentum: $(J_{0}+\Bar{J}_{0}) \vert (p,w) \rangle \rangle_{N} = \Hat{p} \vert (p,w) \rangle \rangle_{N} =0$. The two types of Ishibashi states are 
\begin{equation}
     \vert (p,0) \rangle \rangle_{D} = \exp{  \Big\{  - \sum_{n=1}^{\infty}  \frac{1}{n} \alpha_{-n} \Bar{\alpha}_{-n}   \Big\} } \vert (p,0) \rangle ,
\end{equation}
\begin{equation}
  \vert (0,w) \rangle \rangle_{N} = \exp{ \Big\{ \sum_{n=1}^{\infty}  \frac{1}{n}\alpha_{-n} \Bar{\alpha}_{-n}   \Big\} } \vert (0,w) \rangle .
\end{equation}

\paragraph{$\widehat{su(2)}_{k}  $ Ishibashi state expansion at level-$0$: }

Non-Abelian Kac-Moody Ishibashi state expansions are complicated in general. In this work, we only perform the level-$0$ expansion of $\widehat{su(2)}_{k}  $ Ishibashi state to show the single poles in expansion coefficients when the corresponding $\widehat{su(2)}_{k}$ module is \textbf{integrable}. Consider the $\widehat{su(2)}_{k}$ current algebra in the spin basis, whose commutation relations are  
\begin{equation}
    [\hat{j}_{m}^{3},\hat{j}_{n}^{3}]=\frac{m k}{2} \delta_{m+n,0}, \quad  [\hat{j}_{m}^{3}, \hat{j}_{n}^{\pm} ] = \pm \hat{j}_{m+n}^{\pm} , \quad  [ \hat{j}_{m}^{+} , \hat{j}_{n}^{-}]=km \delta_{m+n,0} +2\hat{j}_{m+n}^{3}. 
\end{equation}
The $\widehat{su(2)}_{k}$ primary states $\vert v_{h,q} \rangle$ are labeled by a pair of numbers $(h,q)$ and satisfy
\begin{equation}
    \hat{j}_{n}^{3} \vert v_{h,q} \rangle = \hat{j}_{n}^{\pm} \vert v_{h,q} \rangle =0, \quad \forall n \in \mathbb{Z}^{+} ,  \quad  \hat{j}_{0}^{3}  \vert v_{h,q} \rangle =  \frac{q}{2}  \vert v_{h,q} \rangle , \quad  \hat{j}_{0}^{+} \vert v_{h,q} \rangle =  0.
\end{equation} 
The level-$0$ vectors in the $\widehat{su(2)}_{k}$ module $V_{h,q}$ are $(j_{0}^{-})^{m} \vert v_{h,q} \rangle $, $m \in \mathbb{N}$. From the $\widehat{su(2)}_{k}$ current algebra, we can show that all level-$0$ vectors are orthogonal. Hence, we only need to compute the norm of $(j_{0}^{-})^{m} \vert v_{h,q} \rangle $ to determine the level-$0$ expansion of $\vert V_{h,q} \rangle \rangle_{\omega}$
\begin{equation*}
   \langle v_{h,q} \vert (j_{0}^{+})^{m} (j_{0}^{-})^{m} \vert v_{h,q} \rangle =m!(q-m+1)_{m},
\end{equation*}
\begin{equation}
    \sum_{m=0}^{\infty} \frac{1}{m!(q-m+1)_{m}} \vert (j_{0}^{-})^{m}  v_{h,q}  \otimes V_{\omega} U (\Bar{j}_{0}^{-})^{m} \Bar{v}_{h,q}   \rangle.
\end{equation}
When $q \in \mathbb{N} $, that is when the module $V_{h,q}$ is integrable, $ (j_{0}^{-})^{q+1} \vert v_{h,q}\rangle$ is a level-$0$ null vector. The expansion coefficients of terms $ (j_{0}^{-})^{m} \vert v_{h,q}\rangle$ with $m \ge (q+1)$, process single poles from $(\langle v_{h,q} \vert (j_{0}^{+})^{m} (j_{0}^{-})^{m} \vert v_{h,q}\rangle)^{-1}$.

\

\section{Free field realizations to Ishibashi states}
\label{sec: fre ap}

From this discussion in the previous section, we have already seen that for free bosonic CFT$_2$, expanding the Ishibashi states in terms of the asymptotic states is simple. This is true for other free field CFT$_2$, such as the free fermion and ghost field theories. Hence, we apply the free field realization to the Ishibashi states. We claim that a CFT$_2$ admits a free field realization if \cite{Bouwknegt:1990wa}
\begin{itemize}
    \item  The chiral algebras $\mathcal{A}$ and $\Bar{\mathcal{A}}$ can be realized using the free fields. 

    \item  The existence of projection maps from free field Fock space modules to the irreducible chiral modules in the CFT$_2$. This is called the Fock space resolution of $\mathcal{A}$ and $\Bar{\mathcal{A}}$ modules.

    \item  Correlation functions of the CFT$_2$ can be calculated using the free field vertex operators. 
\end{itemize}
The third requirement can be relaxed if we are only interested in expanding Ishibashi states in asymptotic states. However, our ultimate objective is to calculate the correlation functions $\mathcal{F}_{g,n,b,c}^{x}$. Hence, the realizations with all three conditions satisfied are preferred.

A prototype of free field realization is the Coulomb gas formalism of both unitary and non-unitary Virasoro minimal models $\mathcal{M}(p,p')$  \cite{Dotsenko:1984nm, Dotsenko:1984ad, Dotsenko:1985hi, Felder:1988zp, Felder:1989ve}. In the Coulomb gas formalism, the Virasoro algebra with central charge $c_{p,p'}$ is realized by a background-charged bosonic theory. The irreducible representations $\mathcal{H}_{r,s}^{p,p'}$ in $\mathcal{M}(p,p')$ are isomorphic to the zeroth cohomology space of a complex of Fock spaces $C_{r,s}^{p,p'}$ \cite{Felder:1988zp}. The computations of chiral primary correlation functions on the sphere are achieved using bosonic vertex operators, with extra insertions of the so-called screening operators \cite{Dotsenko:1984nm, Dotsenko:1984ad, Dotsenko:1985hi}. To apply free field realization to Ishibashi states, our focus is on the Fock space resolutions of the chiral modules, which have been later extended to various models \cite{Bernard:1989iy, Bouwknegt:1989xa, Bouwknegt:1989jf, Jayaraman:1989tu, Distler:1989xv, Bouwknegt:1990fb, Bouwknegt:1990wa}. Some conventional works have already applied Fock space resolutions to Ishibashi states with various underlying motivations \cite{Kawai:2002vd, Parkhomenko:2003gy, Caldeira:2003zz, Hemming:2004dm, Parkhomenko:2004ab}. This section reviews the application of Ishibashi states in Virasoro minimal models $\mathcal{M}(p,p')$ and extends it to $N=1$ minimal models $\mathcal{SM}(p,p')$, whose structures are analogous to $\mathcal{M}(p,p')$. We also give a free fermion expression of the Ising-model Ishibashi states. More applications of free field resolutions to more generic Ishibashi states are working in progress. 

\

\subsection{Free boson realization of Virasoro minimal Ishibashi states} \label{subsec: bos and fer}

First, we review diagonal Virasoro minimal models $\mathcal{M}(p,p')$, the Coulomb gas formalism of $\mathcal{M}(p,p')$, and the application of the Coulomb gas formalism to Virasoro minimal model Ishibashi states. We give a compactified version of the Coulomb gas formalism of Ishibashi states.

Virasoro minimal models $\mathcal{M}(p,p')$ are rational CFT$_2$, whose Hilbert spaces $\mathcal{H}$ are finite direct sum over irreducible representations of Virasoro conformal algebra $\mathfrak{Vir} \otimes \mathfrak{Vir}$. We only consider Virasoro minimal models with diagonal modular invariants in this work. The finiteness of the direct sum in the Hilbert space $\mathcal{H}$ leads to the following rational condition
\begin{equation}
    p' \alpha_{+} + p \alpha_{-}=0 , \label{eq: rat con min}
\end{equation}
where $p > p' \ge 2 $ are two coprime positive integers. The solutions to the rational conditions give the following values $c_{p,p'}$ of Virasoro central charges
\begin{equation}
    c_{p,p'} = 1- \frac{6(p-p')^2}{pp'} . \label{eq: Vir cen min}
\end{equation}

The primary conformal weights of minimal model degenerate modules $V_{r,s}^{p,p'}$ have the following property
\begin{equation}
    h_{r,s}^{p,p'}  =h_{p'-r,p-s}^{p,p'},
\end{equation}
indicating the existence of two distinct singular vectors at level-$rs$ and level-$(p'-r)(p-s)$ respectively. Further, the two submodules generated by two singular vectors are also degenerate modules containing two singular vectors
\begin{equation}
    h_{r,s}^{p,p'} +rs = h_{p'+r,p-s}^{p,p'} = h_{p'-r,p+s}^{p,p'} , \quad  h_{r,s}^{p,p'}+(p'-r)(p-s) =  h_{r,2p-s}^{p,p'} = h_{2p'-r, s}^{p,p'} . \label{eq: h tow min mod} 
\end{equation}
We can verify that $V_{p'+r,p-s}^{p,p'}$ and $V_{r,2p-s}^{p,p'}$ share the same two singular vectors.

\begin{figure} [htbp]
    \centering
     \begin{tikzpicture}
     \filldraw [black] (0,2.5) circle (0.1)
     (-1,2)  circle (0.1)
     (1,2)  circle (0.1)
     (-1,1)  circle (0.1)
     (1,1)  circle (0.1)
     (-1,0)  circle (0.1)
     (1,0)  circle (0.1)
     (-1,-1)  circle (0.1)
     (1,-1)  circle (0.1)
     (-1,-2)  circle (0.1)
     (1,-2)  circle (0.1); 
     \filldraw [gray] (0,0.7) circle (0.04)
     (0,0.5) circle (0.04)
     (0,0.3) circle (0.04);
     \draw (0, 2.9) node {$(r,s)$}
     (-2,2) node {$(r,2p-s)$} 
     (2.23,2) node {$(p'+r,p-s)$}
     (-2,1) node {$(r,2p+s)$} 
     (2,1) node {$(2p'+r,s)$}
     (-4,-1) node {$(r,kp+ (-1)^{k} s + p[1-(-1)^{k}]/2  )$} 
     (4,-1) node {$(r+kp',(-1)^{k} s + p[1-(-1)^{k}]/2  )$};
     \draw [very thick, gray, ->]  (-0.1,2.4) -- (-0.9,2.1) ;
    \draw [very thick, gray, ->] (0.1,2.4) -- (0.9,2.1) ; 
    \draw [very thick, gray, ->] (1,1.85) -- (1,1.15) ;
    \draw [very thick, gray, ->] (-1,1.85) -- (-1,1.15) ;
    \draw [very thick, gray, ->] (-0.9,1.85) -- (0.9,1.15) ;
    \draw [very thick, gray, ->] (0.9,1.85) -- (-0.9,1.15) ;
    \draw [very thick, gray, ->] (1,-0.15) -- (1,-0.85) ;
    \draw [very thick, gray, ->] (-1,-0.15) -- (-1,-0.85) ;
    \draw [very thick, gray, ->] (1,-1.15) -- (1,-1.85) ;
    \draw [very thick, gray, ->] (-1,-1.15) -- (-1,-1.85) ;
    \draw [very thick, gray, ->] (-0.9,-0.15) -- (0.9,-0.85) ;
    \draw [very thick, gray, ->] (0.9,-0.15) -- (-0.9,-0.85) ;
    \draw [very thick, gray, ->] (-0.9,-1.15) -- (0.9,-1.85) ;
    \draw [very thick, gray, ->] (0.9,-1.15) -- (-0.9,-1.85) ;
     \end{tikzpicture}
    \caption{The two infinite classes of null vectors of the degenerate Verma module $V_{r,s}^{p,p'}$ in the Virasoro minimal model $\mathcal{M}(p,p')$. The same structure also applies to degenerate modules in $N=1$ minimal models $\mathcal{SM}(p,p')$}
    \label{fig: min null doub}
\end{figure}
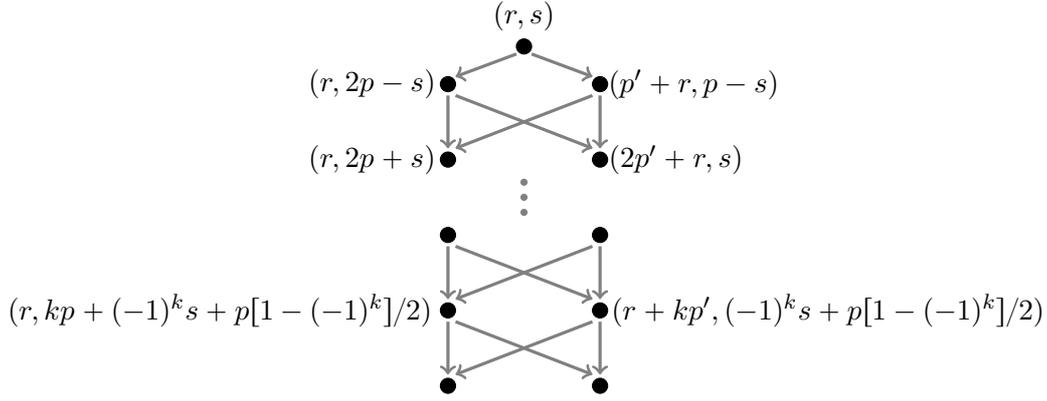

This procedure can be iterated infinite times, with two singular sub-modules appearing at each step. Hence, the Verma modules in minimal models contain two infinite classes of singular vectors, as shown in Figure \ref{fig: min null doub}. From this singular vector structure, we can express the minimal irreducible characters $\chi_{r,s}^{p,p'}(q)$ 
\begin{equation}
    \chi_{r,s}^{p,p'}(q)= \frac{q^{\frac{1-c_{Vir}}{24}}}{\eta(q)}\Big[ q^{h_{r,s}} + \sum_{k=1}^{\infty} (-1)^{k} (q^{h_{r+k'p, (-1)^{k} s + [1-(-1)^{k}] \frac{p}{2} }}  + q^{h_{r, kp+ (-1)^{k} s + [1-(-1)^{k}] \frac{p}{2} }}) \Big],  \label{eq: min ch}
\end{equation}
where $\eta(q):= q^{\frac{1}{24}} \prod_{n=1}^{\infty} (1-q^{n}) $ is the Dedekind $\eta$-function. An important alternative expression of chiral characters in Virasoro minimal models is the so-called $K$-function expression since they give the hint of Fock space resolution
\begin{equation}
    \chi_{r,s}^{p,p'}( q) = K_{r,s}^{p,p'}(q) - K_{r,-s}^{p,p'}(q) , \label{eq: K fun exp ch}
\end{equation}
where
\begin{equation}
     K_{r, \pm s}^{p,p'}(q) := \frac{1}{\eta(q)} \sum_{n \in \mathbb{Z}}  q^{\frac{(2pp' n + \lambda_{r, \pm s}  )^{2}}{4pp'}} , \quad \lambda_{r, \pm s} := pr  \mp  p's.
\end{equation}

\paragraph{Fock space resolution of chiral irreducible modules $\mathcal{H}_{r,s}^{p,p'}$:} 

A single free boson CFT$_2$ has Virasoro central charge $c_{Vir}=1$. To match the central charges $c_{p,p'}$, we place a background charge on the sphere, which will modify the energy-stress tensor to
\begin{equation}
     T(z) = - \frac{1}{2} :   \partial \phi  (z) \partial \phi (z) :  + i \sqrt{2} \alpha_{0} \partial^{2} \phi(w) , \quad  2\alpha_{0}= \alpha_{+} + \alpha_{-} = \frac{p-p'}{\sqrt{pp'}} ,
\end{equation}
which realizes the Virasoro algebra with $c_{p,p'}$
\begin{equation}
    c_{Vir}= 1-24\alpha_{0}^{2} = c_{p,p'} .
\end{equation}

Next, we discuss the Fock space resolution to $\mathcal{H}_{r,s}^{p,p'}$. The background-charged Fock space primary vectors are obtained by acting chiral vertex operators on the vacuum vector
\begin{equation}
     \vert \alpha ,\alpha_{0} \rangle = V_{\alpha,\alpha_{0}} (0) \vert 0 \rangle =  : \exp{ (i\sqrt{2}  \alpha  \phi )  } : (0) \vert 0 \rangle  ,
\end{equation}
and they satisfy 
\begin{equation*}
    a_{0} \vert \alpha ,\alpha_{0} \rangle = \sqrt{2}\alpha  \vert \alpha ,\alpha_{0} \rangle , \quad a_{n}  \vert \alpha ,\alpha_{0} \rangle   = 0 ,\quad \forall n >0 ,
\end{equation*}
\begin{equation}
   L_{0} \vert \alpha ,\alpha_{0} \rangle =  \big( \alpha^{2} -2\alpha_{0}\alpha \big) \vert \alpha ,\alpha_{0} \rangle , \quad L_{n}  \vert \alpha ,\alpha_{0} \rangle =0 , \quad \forall n >0 ,
\end{equation}
where $a_{n}$ are the Laurant modes of $i \partial \phi (z)$
\begin{equation}
    i \partial \phi (z)= \sum_{n \in \mathbb{Z} } \frac{a_{n}}{z^{n+1}}  .
\end{equation}
The Virasoro generators of background-charged boson $\phi$ are given by
\begin{equation}
    L_{n}= \frac{1}{2}\sum_{k\in \mathbb{Z} }  : a_{n-k} a_{k} : -\sqrt{2} \alpha_{0} (n+1) a_{n} , \quad  L_{0} = \sum_{k \in \mathbb{Z}^{+} } a_{-k}a_{k} +\frac{1}{2} a_{0}^{2} -\sqrt{2} \alpha_{0} a_{0} .
\end{equation}
The character of the Fock space $F_{\alpha,\alpha_{0}}$ is
\begin{equation}
    \chi_{\alpha, \alpha_{0}}(q) : =  \text{Tr}_{F_{\alpha,\alpha_{0}}} \: q^{L_{0} -\frac{c_{Vir}}{24} }  =  \frac{q^{(\alpha-\alpha_{0})^{2}}}{\eta(q)}  .
\end{equation}
The $U(1)$ charges carried by the Fock space primaries that have the conformal weights $h_{r,s}^{p,p'}$ are 
\begin{equation}
    \alpha_{r,s}= \frac{1}{2}(1-r)\alpha_{+} + \frac{1}{2}(1-s)\alpha_{-} .
\end{equation}
The primary vectors of those charged Fock spaces and their primary vectors are denoted by $F_{r,s}^{p,p'}$ and $\vert  \alpha_{r,s} \rangle$ respectively.

\begin{figure} [htbp]
    \centering
     \begin{tikzpicture} 

\draw (0, 3) node {$F_{r,s}$} ;
     
     \filldraw [black] (0,2.5) circle (0.1)
     (-1,2)  circle (0.1)
     (1,2)  circle (0.1)
     (-1,1)  circle (0.1)
     (1,1)  circle (0.1)
     (-1,0)  circle (0.1)
     (1,0)  circle (0.1)
     (-1,-1)  circle (0.1)
     (1,-1)  circle (0.1)
     (-1,-2)  circle (0.1)
     (1,-2)  circle (0.1); 
      \filldraw [gray] (0,-2.3) circle (0.04)
     (0,-2.5) circle (0.04)
     (0,-2.7) circle (0.04);

     \draw [very thick, gray, ->] (1,0.15) -- (1,0.85) ;
    \draw [very thick, gray, ->] (-1,0.85) -- (-1,0.15) ;
     \draw [very thick, gray, ->]  (-0.1,2.4) -- (-0.9,2.1) ;

     \draw [very thick, gray, ->] (0.9,0.85) -- (-0.9,0.15) ;
    \draw [very thick, gray, ->] (0.9,0.15) -- (-0.9,0.85) ;
    
    \draw [very thick, gray, ->] (0.9,2.1) -- (0.1,2.4) ; 
    \draw [very thick, gray, ->] (1,1.85) -- (1,1.15) ;
    \draw [very thick, gray, ->] (-1,1.15) -- (-1,1.85) ;
    \draw [very thick, gray, ->] (0.9,1.85) -- (-0.9,1.15) ;
    \draw [very thick, gray, ->] (0.9,1.15) -- (-0.9,1.85) ;
    \draw [very thick, gray, ->] (1,-0.15) -- (1,-0.85) ;
    \draw [very thick, gray, ->] (-1,-0.85) -- (-1,-0.15) ;
    \draw [very thick, gray, ->] (1,-1.85) -- (1,-1.15) ;
    \draw [very thick, gray, ->] (-1,-1.15) -- (-1,-1.85) ;
    \draw [very thick, gray, ->] (0.9,-0.15) -- (-0.9,-0.85) ;
    \draw [very thick, gray, ->] (0.9,-0.85) -- (-0.9,-0.15) ;
    \draw [very thick, gray, ->] (0.9,-1.15) -- (-0.9,-1.85) ;
    \draw [very thick, gray, ->] (0.9,-1.85) -- (-0.9,-1.15) ;

    \draw [very thick, red, ->] (-3,2) -- (-2,2) ; 
    \draw [very thick, red, ->] (-3,1) -- (-2,1) ; 
    \draw [very thick, red, ->] (-3,0) -- (-2,0) ; 
    \draw [very thick, red, ->] (-3,-1) -- (-2,-1) ; 
    \draw [very thick, red, ->] (-3,-2) -- (-2,-2) ; 

    \draw [very thick, red, ->] (2,2) -- (3,2) ; 
    \draw [very thick, red, ->] (2,1) -- (3,1) ; 
    \draw [very thick, red, ->] (2,0) -- (3,0) ; 
    \draw [very thick, red, ->] (2,-1) -- (3,-1) ; 
    \draw [very thick, red, ->] (2,-2) -- (3,-2) ; 

    \draw (5,2.5) node {$F_{r,-s}$} ;

     \filldraw [black] (5,2) circle (0.1)
     (4,1)  circle (0.1)
     (6,1)  circle (0.1)
     (4,0)  circle (0.1)
     (6,0)  circle (0.1)
     (4,-1)  circle (0.1)
     (6,-1)  circle (0.1)
     (4,-2)  circle (0.1)
     (6,-2)  circle (0.1); 
      \filldraw [gray] (5,-2.3) circle (0.04)
     (5,-2.5) circle (0.04)
     (5,-2.7) circle (0.04);

     \draw [very thick, gray, ->] (4.9,1.9) -- (4.1,1.1) ;
     \draw [very thick, gray, ->] (5.9,1.1) -- (5.1,1.9) ;
     \draw [very thick, gray, ->] (6,0.85) -- (6,0.15) ; 
     \draw [very thick, gray, ->] (4,0.15) -- (4,0.85) ; 
     \draw [very thick, gray, ->] (4,-0.15) -- (4,-0.85) ;
     \draw [very thick, gray, ->] (6,-0.85) -- (6,-0.15) ; 
     \draw [very thick, gray, ->] (6,-1.15) -- (6,-1.85) ; 
     \draw [very thick, gray, ->] (4,-1.85) -- (4,-1.15) ; 

     \draw [very thick, gray, ->] (5.9,-1.85) -- (4.1,-1.15) ; 
     \draw [very thick, gray, ->] (5.9,-1.15) -- (4.1,-1.85) ; 

     \draw [very thick, gray, ->] (5.9,0.85) -- (4.1,0.15) ; 
     \draw [very thick, gray, ->] (5.9,0.15) -- (4.1,0.85) ; 

     \draw [very thick, gray, ->] (5.9,-0.85) -- (4.1,-0.15) ; 
     \draw [very thick, gray, ->] (5.9,-0.15) -- (4.1,-0.85) ;

     \draw (-5,2.5) node {$F_{r,2p-s}$} ;

     \filldraw [black] (-5,2) circle (0.1)
     (-4,1)  circle (0.1)
     (-6,1)  circle (0.1)
     (-4,0)  circle (0.1)
     (-6,0)  circle (0.1)
     (-4,-1)  circle (0.1)
     (-6,-1)  circle (0.1)
     (-4,-2)  circle (0.1)
     (-6,-2)  circle (0.1); 
      \filldraw [gray] (-5,-2.3) circle (0.04)
     (-5,-2.5) circle (0.04)
     (-5,-2.7) circle (0.04);

     \draw [very thick, gray, ->] (-5.1,1.9) -- (-5.9,1.1) ;
     \draw [very thick, gray, ->] (-4.1,1.1) -- (-4.9,1.9) ;
     \draw [very thick, gray, ->] (-4,0.85) -- (-4,0.15) ; 
     \draw [very thick, gray, ->] (-6,0.15) -- (-6,0.85) ; 
     \draw [very thick, gray, ->] (-6,-0.15) -- (-6,-0.85) ;
     \draw [very thick, gray, ->] (-4,-0.85) -- (-4,-0.15) ; 
     \draw [very thick, gray, ->] (-4,-1.15) -- (-4,-1.85) ; 
     \draw [very thick, gray, ->] (-6,-1.85) -- (-6,-1.15) ; 

     \draw [very thick, gray, ->] (-4.1,-1.85) -- (-5.9,-1.15) ; 
     \draw [very thick, gray, ->] (-4.1,-1.15) -- (-5.9,-1.85) ; 

     \draw [very thick, gray, ->] (-4.1,0.85) -- (-5.9,0.15) ; 
     \draw [very thick, gray, ->] (-4.1,0.15) -- (-5.9,0.85) ; 

     \draw [very thick, gray, ->] (-4.1,-0.85) -- (-5.9,-0.15) ; 
     \draw [very thick, gray, ->] (-4.1,-0.15) -- (-5.9,-0.85) ;
    
     \end{tikzpicture}
    \caption{The red horizontal arrows denote the action of BRST charges $Q_{-}^{s}$ and $Q_{-}^{p-s}$ on all the primary vectors in the BRST complex. Only the cohomology space of the center Fock space $F_{r,s}$ is non-trivial and is isomorphic to the irreducible representation $\mathcal{H}_{r,s}^{p,p'}$.} 
    \label{fig: BRST compl}
\end{figure}
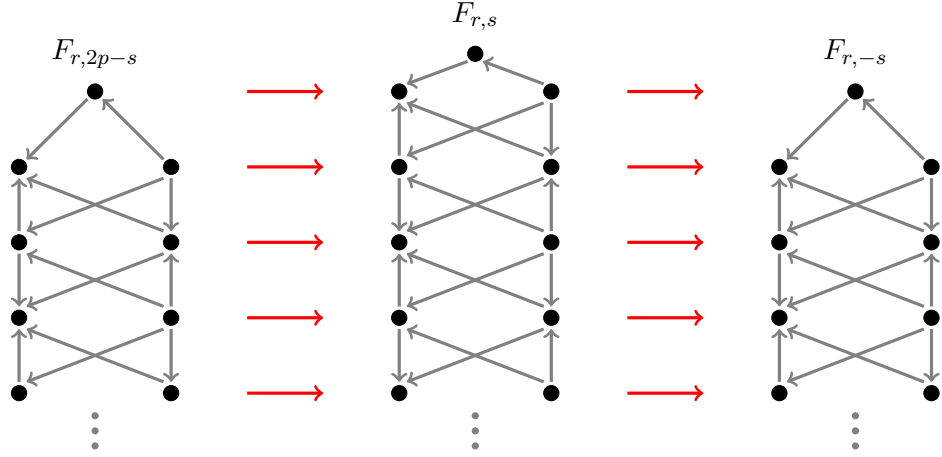

We need to find out the projection from the background charged Fock spaces $F_{r,s}^{p,p'}$ to irreducible representations $\mathcal{H}_{r,s}^{p,p'}$. The hint is hidden in the $K$-function expression (\ref{eq: K fun exp ch}). The terms in $K_{r, \pm s}^{p,p'}$ can be viewed as the characters of background-charged Fock spaces $F_{r, \pm s -2 j p }$, $j\in \mathbb{Z}$ respectively. We need to find the mappings that change the $U(1)$ charge but keep the conformal properties to form a complex of the spaces $F_{r,\pm s - 2 j p}$. Those actions are Virasoro intertwiners, which are constructed from the so-called screening operators $Q_{\pm}$.  $Q_{\pm}$ are non-local operators constructed from a contour integral over vertex operators with conformal weight $h=1$ \cite{Dotsenko:1984nm}
\begin{equation}
    Q_{\pm} := \oint\frac{dz}{2\pi i} V_{\pm}(z) =  \oint\frac{dz}{2\pi i} e^{i\sqrt{2} \alpha_{\pm}  \phi } (z), \quad [Q_{\pm}, L_{k}] =0 ,\quad \forall k\in\mathbb{Z} .
\end{equation}
The Virasoro intertwiners are constructed from radial ordered multiple contour integrals of $V_{\pm}$ \cite{Felder:1988zp}
\begin{equation}
    Q_{\pm}^{m} : = \frac{1}{m} \prod_{j=1}^{m} \oint \frac{dz_{j}}{2\pi i}  V_{\pm} (z_{j})  , \quad  1 > \vert z_{1} \vert  > \vert z_{2} \vert > \cdots > \vert z_{m} \vert >0 . \label{eq: def BRST op Vir}
\end{equation}
We use intertwiners $Q_{-}^{s}$ and $Q_{-}^{p-s}$ to construct a complex $C_{r,s}^{p,p'}$ of the Fock spaces $F_{r, -s \pm 2 j p }^{p,p'}$
\begin{equation}
    C_{r,s}^{p,p'}: \quad \cdots \to   F_{r,2p+s}^{p,p'}  \overset{Q_{-}^{s}}{ \longrightarrow}  F_{r,2p-s}^{p,p'}  \overset{Q_{-}^{p-s}}{ \longrightarrow}  F_{r,s}^{p,p'}  \overset{Q_{-}^{s}}{ \longrightarrow} F_{r, -s}^{p,p'}   \overset{Q_{-}^{p-s}}{ \longrightarrow}  F_{r,s-2p}^{p,p'} \to \cdots .
\end{equation}
The nilpotency of the intertwiners $Q_{-}^{s} Q_{-}^{p-s} =0 $ is proven  \cite{Felder:1988zp}. The explicit structure is shown in Figure \ref{fig: BRST compl} \cite{Felder:1988zp}, from which we conclude that only the zeroth cohomology space (the cohomology of the center Fock space $F_{r,s}^{p,p'}$) is non-trivial, and it is isomorphic to the irreducible representations $\mathcal{H}_{r,s}^{p,p'}$  \cite{Felder:1988zp}
\begin{equation}
    H^{j} (C_{r,s}^{p,p'}) = \begin{cases}
        0 &  j \ne 0 ;  \\  \frac{\text{Ker} Q_{-}^{s}}{\text{Im} Q_{-}^{p-s}}    & j= 0  , 
    \end{cases}  \quad    H^{0} (C_{r,s}^{p,p'}) \cong  \mathcal{H}_{r,s}^{p,p'}.   \label{eq: 0 coh rs}
\end{equation}
(\ref{eq: 0 coh rs}) indicates that taking the trace over the zeroth cohomology space of $C_{r,s}^{p,p'}$ reproduces the $K$-function expression of the minimal model chiral characters 
\begin{equation*}
    \text{Tr}_{H^{0} (C_{r,s}^{p,p'}) } (q^{L_{0}- \frac{c_{Vir}}{24} }) =  \sum_{j \in \mathbb{Z}} \text{Tr}_{ F_{r,s-2jp} } (q^{L_{0}- \frac{c_{Vir}}{24} }) -\sum_{j \in \mathbb{Z}} \text{Tr}_{F_{r,-s-2jp} } (q^{L_{0}- \frac{c_{Vir}}{24} }) 
\end{equation*}
\begin{equation}
    = K_{r,s}^{p,p'}(q) - K_{r,-s}^{p,p'}(q)  = \chi_{r,s}^{p,p'}(q) . 
\end{equation}

\paragraph{The application of Coulomb gas formalism to Ishibashi states:}

Now, we apply (\ref{eq: 0 coh rs}) to Ishibashi states in diagonal Virasoro minimal models $\vert \mathcal{H}_{r,s}^{p,p'} \rangle \rangle \in \mathcal{H}_{r,s}^{p,p'} \otimes U  \mathcal{H}_{r,s}^{p,p'} $, following \cite{Kawai:2002vd, Kawai:2002pz}. We also give a compactified version of this application.

We begin showing the constraint condition on the bosonic Ishibashi states $ \vert \alpha, \Bar{\alpha} , \alpha_{0} \rangle \rangle_{\omega}$. The forms of bosonic Ishibashi states are as follows 
\begin{equation}
    \vert \alpha, \Bar{\alpha} , \alpha_{0} \rangle \rangle_{\omega} =  \exp{  \Big[ \sum_{n=1}^{\infty}  -\frac{\Omega}{n} a_{-n} \Bar{a}_{-n} \Big]  }  \vert \alpha,  \Bar{\alpha} ,\alpha_{0} \rangle .
\end{equation}
The $(L_{0}-\Bar{L}_{0}) \vert \alpha , \alpha_{0} \rangle \rangle_{\omega} $ condition requires 
\begin{equation}
    (\alpha-\Bar{\alpha}) (\alpha+\Bar{\alpha}-2\alpha_{0}) \vert \alpha, \Bar{\alpha} , \alpha_{0} \rangle \rangle_{\omega}=0,
\end{equation}
and the $(L_{n}-\Bar{L}_{-n}) \vert \alpha, \Bar{\alpha} , \alpha_{0} \rangle \rangle_{\omega}$ condition requires
\begin{equation*}
    (L_{n}-\Bar{L}_{-n}) \vert \alpha, \Bar{\alpha} , \alpha_{0} \rangle \rangle_{\omega} = 
\end{equation*}
\begin{equation}
   \Big\{ \sqrt{2}  \Bar{a}_{-n}  \big[ (\Omega-1) n \alpha_{0} +(\Omega+1) \alpha_{0} - \Omega \alpha - \Bar{\alpha}  \big]  + \frac{1}{2} (\Omega^{2}-1) \Bar{a}_{j}  \Bar{a}_{j-n} \Big\} \vert \alpha ,  \Bar{\alpha} , \alpha_{0} \rangle \rangle_{\omega} =0.
\end{equation}
The Virasoro gluing conditions are satisfied only when 
\begin{equation}
    \Omega=1, \quad  (\alpha+\Bar{\alpha}-2\alpha_{0}) = 0 .
\end{equation}
Hence, the bosonic Ishibashi states we use to express the minimal model Ishibashi states $\vert \mathcal{H}_{r,s}^{p,p'} \rangle \rangle$ are
\begin{equation}
    \vert \alpha_{r,s},  2\alpha_{0} - \alpha_{r,s}  ,  \alpha_{0}  \rangle \rangle_{N} .
\end{equation}

The expression of $\vert \mathcal{H}_{r,s}^{p,p'}  \rangle \rangle$ is
\begin{equation}
   \vert \mathcal{H}_{r,s}^{p,p'}  \rangle \rangle   =  \sum_{j\in \mathbb{Z}} \kappa_{\pm; j} \vert \alpha_{r, \pm s -2jp} ,  2\alpha_{0} - \alpha_{r, \pm s -2jp } ,  \alpha_{0}  \rangle \rangle_{N} ,  \label{eq: Cou gas Ishi}
\end{equation}
where $ \kappa_{\pm;j} \in U(1)$ are some undetermined phases. We write the dual Ishibashi states as
\begin{equation}
    \langle  \langle \mathcal{H}_{r,s}^{p,p'} \vert   = \sum_{n}  \kappa_{\pm; j}'  \:  _{N}\langle \langle \alpha_{r, \pm s  -2jp }  ,  2\alpha_{0} - \alpha_{r, \pm s -2jp } ,  \alpha_{0}  \vert ,
\end{equation}
where $\kappa_{\pm; j}' \kappa_{\pm; j} = \pm  1$. The overlaps reproduce the $K$-function expression of the irreducible minimal chiral characters
\begin{equation}
  \langle \langle \mathcal{H}_{r,s}^{p,p'} \vert q^{ (L_{0}+\Bar{L}_{0}- \frac{c_{Vir}}{12})}   \vert \mathcal{H}_{r,s}^{p,p'} \rangle \rangle =  K_{r,s}^{p,p'}(q^{2}) -K_{r,-s}^{p,p'}(q^{2}) = \chi_{r,s}^{p,p'}(q^{2}) .
\end{equation}
The orthogonal property of the overlaps between different Virasoro Ishibashi states is ensured by the fact that different complexes consist of different Fock spaces and the orthogonal property of bosonic Ishibashi states.

\paragraph{Compactified version of the Coulomb gas formalisms:}

Now, we consider a compactified version of the Fock space module resolutions of $\mathcal{H}_{r,s}^{p,p'}$, and apply it to Ishibashi states $\vert \mathcal{H}_{r,s}^{p,p'} \rangle \rangle$.

Consider a compactified background-charged bosonic field whose target space is a circle with radius $R$. The compact bosonic vertex operators are labeled by the momentum $k$ and the winding number $w$ 
\begin{equation}
   V_{k,w}(z,\bar{z}) = \exp{ \Big[  i \sqrt{2} (\frac{k}{R} + \frac{wR}{2}) \phi + i \sqrt{2} (\frac{k}{R} - \frac{wR}{2})  \Bar{\phi}  \Big] } (z,\Bar{z}) .
\end{equation}
The background-charged Fock spaces $F_{(k,w)} $ primaries are obtained by
\begin{equation}
    \vert v_{k,w}\rangle = \exp{ \Big[  i \sqrt{2} (\frac{k}{R} + \frac{wR}{2}) \phi + i \sqrt{2} (\frac{k}{R} - \frac{wR}{2})  \Bar{\phi}  \Big] } \vert 0 \rangle ,
\end{equation}
whose conformal weights are
\begin{equation}
    (h,\Bar{h}) = \Big[   \big( \frac{k}{R} + \frac{wR}{2}  \big) \big( \frac{k}{R} + \frac{wR}{2} -2\alpha_{0} \big)  ,   \big( \frac{k}{R} - \frac{wR}{2}  \big) \big( \frac{k}{R} - \frac{wR}{2} -2\alpha_{0} \big) \Big] .
\end{equation}

We begin with the constraint conditions on the bosonic Ishibashi states. The $(L_{0}-\Bar{L}_{0})\vert k,w ,\alpha_{0} \rangle \rangle_{\omega} =0 $ condition requires 
\begin{equation}
    w(k-\alpha_{0} R)=0 .
\end{equation}
The $(L_{n}-\Bar{L}_{-n})\vert k,w ,\alpha_{0} \rangle \rangle_{\omega} =0  $ condition requires 
\begin{equation*}
    \Big\{ \sqrt{2}  \Bar{a}_{-n}  \big[ (\Omega-1) n \alpha_{0} +(\Omega+1) \alpha_{0} - \Omega (\frac{k}{R} + \frac{wR}{2}) -  (\frac{k}{R} -\frac{wR}{2})  \big] 
\end{equation*}
\begin{equation}
    + \frac{1}{2} (\Omega^{2}-1) \Bar{a}_{j}  \Bar{a}_{j-n}  \Big\} \vert k, w , \alpha_{0} \rangle \rangle_{\omega} =0 .
\end{equation}
Hence, the bosonic Ishibashi states allowed to express $\vert \mathcal{H}_{r,s}^{p,p'} \rangle \rangle$ are those with
\begin{equation}
    \Omega=1 , \quad  k= \alpha_{0} R.
\end{equation}

The screening compactified version of operators $Q_{w_{\pm}}$ are 
\begin{equation}
    Q_{w_{\pm}} = \oint \frac{dz}{2\pi i}  V_{ k=\alpha_{0} R , w_{\pm} } , \quad w_{\pm}= \frac{2}{R} ( \alpha_{\pm} -\alpha_{0} ) .
\end{equation}
Next, we find the required winding numbers $w$ and radius $R$ to construct the Fock space complex. The results are
\begin{equation}
    w_{r,\pm s -2jp }^{p,p'} =  2 jp p' + \lambda_{r,\pm s}  + p -  p'  , \quad  R= \frac{1}{\sqrt{pp'}}  .
\end{equation}
This solution leads to the fact that the momentum and winding numbers in the screening operates are fractional $w_{\pm} = \pm \frac{p+p'}{pp'}$ and $k=\frac{p'-p}{2pp'}$. We use $Q_{w_{-}}^{s}$ and $Q_{w_{-}}^{p-s}$ to construct a complex $C_{r,s}^{p,p'}$ 
\begin{equation}
    C_{r,s}^{p,p'}: \quad \cdots   \overset{Q_{k_{-}}^{p-s}}{ \longrightarrow}  F_{\frac{p-p'}{2pp'} ,w_{r,s} }  \overset{Q_{k_{-}}^{s}}{ \longrightarrow} F_{ \frac{p-p'}{2pp'} ,w_{r, -s} }   \overset{Q_{k_{-}}^{p-s}}{ \longrightarrow} \cdots ,
\end{equation}
such that its zeroth cohomology space of $C_{r,s}^{p,p'}$ is isomorphic to the irreducible representation $\mathcal{H}_{r,s}^{p,p'}$. Applying this to the Ishibashi states, we write down $\vert \mathcal{H}_{r,s}^{p,p'} \rangle \rangle$ as
\begin{equation}
    \vert \mathcal{H}_{r,s}^{p,p'} \rangle \rangle = \sum_{j\in \mathbb{Z}}   \kappa_{\pm; j}  \vert ( \frac{p-p'}{2pp'} , w_{r,\pm s -2jp}^{p,p'} ,\alpha_{0} )   \rangle \rangle_{N} .
\end{equation}

\

\subsection{Free boson realization of $N=1$ minimal model Ishibashi states}

The $N=1$ minimal models $\mathcal{SM}(p,p')$ are analogous to the Virasoro minimal models $\mathcal{M}(p,p')$. Since $N=1$ minimal models are also RCFT, the finiteness of the chiral modules requires the $N=1$ rational condition
\begin{equation}
    p'\alpha_{+}^{N=1} + p\alpha_{-}^{N=1} =0 .
\end{equation}
The solution of the rational conditions are
\begin{equation}
    \Hat{c}_{p,p'}=1- \frac{2(p-p')^{2}}{pp'}. 
\end{equation}
The primary conformal weights of degenerate $N=1$ modules $V_{r,s}^{N=1}$ in both the NS and R sectors at $\Hat{c}_{p,p'}$ are given by
\begin{equation}
    h_{r,s}^{N=1} (\Hat{c}_{p,p'})  =  \frac{(rp-sp')^{2}   -(p-p')^{2} }{8pp'}  + \frac{1-(-1)^{r-s}}{32}  .
\end{equation}
It is easy to verify that the degenerate $N=1$ modules at $\Hat{c}_{p,p'}$ have the same singular vector structure as the degenerate modules in Virasoro minimal models, as shown in Figure \ref{fig: min null doub}.

The chiral characters of irreducible $N=1$ modules in NS and R sectors are given by \cite{DiFrancesco:1988xz}
\begin{equation}
    \chi^{NS}_{h}= \text{Tr}_{ V_{NS}} q^{L_{0}- \frac{\Hat{c}_{p,p'} }{16}} =   \frac{ \eta(q) }{ \eta( \sqrt{q}) \eta(q^{2})  } q^{h} , 
\end{equation}
\begin{equation}
    \chi^{R}_{h} = \text{Tr}_{ V_{R}} q^{L_{0}- \frac{\Hat{c}_{p,p'} }{16}} =  \sqrt{2} \frac{\eta(q^{2})  }{ [\eta(q)]^{2} } q^{h}  .
\end{equation}
From the results above and the singular vector structure of degenerate modules in $N=1$ modules $\mathcal{H}_{r,s}^{N=1} (\Hat{c}_{p,p'})$, we obtain the explicit expression of chiral characters of irreducible modules in $N=1$ minimal models $\mathcal{H}_{r,s}^{N=1} (\Hat{c}_{p,p'})$. Here, we only write down their $K$-function expressions, which are useful in obtaining the complex for Fock space module resolutions
\begin{equation}
    \chi_{r,s}^{\text{NS}} (\Hat{c}_{p,p'} , q ): =  \text{Tr}_{ \mathcal{H}_{r,s}^{N=1}(\Hat{c}_{p,p'}) } q^{L_{0} - \frac{\Hat{c}_{p,p'}}{16} }  =   \frac{ [\eta(q)]^{2} }{ \eta( \sqrt{q} ) \eta(q^{2})  }   \big[ K_{r,s}^{N=1} (  q) - K_{r,-s}^{N=1} (  q) \big] ,
\end{equation}
\begin{equation}
    \chi_{r,s}^{\text{R}} (\Hat{c}_{p,p'} , q ): =  \text{Tr}_{ \mathcal{H}_{r,s}^{N=1}(\Hat{c}_{p,p'}) } q^{L_{0} - \frac{\Hat{c}_{p,p'}}{16} }  =   \sqrt{2} \frac{\eta(q^{2})  }{ [\eta(q)] }\big[ K_{r,s}^{N=1} ( q) - K_{r,-s}^{N=1} (  q) \big] ,
\end{equation}
where the $N=1$ $K$-functions are 
\begin{equation}
    K_{r, \pm s}^{N=1} (q) := \frac{1}{\eta(q)} \sum_{n  \in \mathbb{Z} } q^{ \frac{( 2pp' n + \lambda_{r, \pm s} )}{8pp'} }  .
\end{equation}

\paragraph{The Coulomb gas formalism of $N=1$ minimal models.}

The Coulomb gas formalism of the minimal models $N=1$ is achieved by adding a background charge term to the sphere with a $N=1$ free CFT$_2$ \cite{Jayaraman:1989as}, which modifies the $N=1$ generators to 
\begin{equation}
    T(z)= - \frac{1}{2} :  \partial \phi   \partial \phi(z)   : - \frac{1}{2} : \psi \partial \psi (z) :  + i  \alpha_{0}^{N=1} \partial^{2} \phi(z)   , \label{eq: N=1 fre T}
\end{equation}
\begin{equation}
    T_{F}(z) =  \frac{i}{2} :  \psi \partial \phi (z) : -  \alpha_{0}^{N=1} \partial \psi (z), \label{eq: N=1 fre G}
\end{equation}
whose $N=1$ central charge is $\Hat{c}=1-8(\alpha_{0}^{N=1})^{2}$. By taking $2\alpha_{0}^{N=1} = \alpha_{+}^{N=1} + \alpha_{-}^{N=1}$, we match the required $N=1$ central charge $\Hat{c}_{p,p'}$.

The background-charged Fock spaces $F_{\alpha, \alpha_{0}}^{N=1}$ of $N=1$ Coulomb gas formalism are separated into NS and R sectors. For NS sector Fock space modules, primary vectors $\vert v_{\alpha,\alpha_{0}}^{N=1} \rangle$ are obtained by acting on the corresponding bosonic vertex operator on the vacuum
\begin{equation}
    \vert v_{\alpha,\alpha_{0}}^{NS , N=1} \rangle =  : e^{i  \alpha \phi } (0): \vert 0 \rangle , \quad L_{0} \vert v_{\alpha,\alpha_{0}}^{NS,N=1} \rangle = \frac{1}{2} \alpha(\alpha -2 \alpha_{0}^{N=1}) \vert v_{\alpha,\alpha_{0}}^{NS, N=1} \rangle .
\end{equation}
For R sector Fock space modules, primary vectors are obtained by
\begin{equation}
    \vert v_{\alpha,\alpha_{0}}^{R , N=1} \rangle = : e^{i  \alpha \phi } (0):   \vert R \rangle , \quad L_{0} \vert v_{\alpha,\alpha_{0}}^{R,N=1} \rangle = \big[ \frac{1}{2} \alpha(\alpha -2 \alpha_{0}^{N=1})  + \frac{1}{16} \big] \vert v_{\alpha,\alpha_{0}}^{R, N=1} \rangle ,
\end{equation}
where $\vert R \rangle$ is the R sector holomorphic ground state with $h=\frac{1}{16}$. $N=1$ Fock space primary vectors with conformal weights $h_{r,s}^{N=1} (\Hat{c}_{p,p'})$ have $U(1)$ charges $\alpha_{r,s}$ in both NS and R sectors
\begin{equation}
    \alpha_{r,s} = \frac{1}{2}  \big[ (1-r)\alpha_{+}^{N=1} + (1-s)\alpha_{-}^{N=1} \big] . 
\end{equation}

The $N=1$ screening operators are defined by contour integrals over the $h=1$ vertex operators. There are bosonic and fermionic $h=1$ operators, we only focus on the fermionic operators 
\begin{equation}
     V_{\pm}^{F}(z) :=  \:  : \psi e^{i \alpha_{\pm}^{N=1}  \phi }  :  (z) , \quad Q_{\pm}^{F} : =  \oint \frac{dz}{2\pi i } V_{\pm}^{F}(z) . 
\end{equation}
We check that whether $Q_{\pm}^{F}$ are $N=1$ intertwiners, the result is that $Q_{\pm}^{F}$ do not anticommute with $G_{r}$ in general
\begin{equation}
    [Q_{\pm}^{F}, L_{n}] =0 , \quad \forall  n  \in \mathbb{Z}.
\end{equation}
\begin{equation}
    \{ G_{r}, Q_{\pm}^{F} \} = (  \alpha_{\pm}^{N=1} -2 \alpha_{0}^{N=1} )  \oint \frac{dw}{2\pi i}  \big(r+\frac{1}{2} \big) w^{r-\frac{1}{2}} : e^{i \alpha_{\pm}^{N=1} \phi }(w)  : = x_{r} , \label{eq: G QF com}
\end{equation}
where $x_{r}$ is one of the Laurent mode of the vertex operator $: e^{i \alpha_{\pm}^{N=1} \phi }(w)  : = \sum_{r} x_{r} / w^{r+\frac{1}{2}} $. Nevertheless, since $: e^{i \alpha_{\pm}^{N=1} \phi }(w)  :$ is a normal ordered product that needs to be regular when $w\to 0$, for $r >-\frac{1}{2} $, $x_{r}=0$, indicating that the actions of $Q_{\pm}^{F}$ still preserve the property of being $N=1$ singular vectors, which is enough in constructing a complex of Fock space modules as a resolution of $N=1$ irreducible modules. The differentials in the complex are defined as
\begin{equation}
    (Q_{\pm}^{N=1})^{(n)} :=  \prod_{i=1}^{n} \oint \frac{dz_{i}}{2\pi i} : \psi e^{ i  \alpha_{\pm}^{N=1}  \phi } (z_{i}) : , 
\end{equation}
where the variables $z_{i}$ are radial ordered as the Virasoro intertwiners (\ref{eq: def BRST op Vir}).

The $K$-function expression of the $N=1$ minimal characters hinted to us to construct the following complex of Fock spaces, for both NS and R sectors
\begin{equation}
   C_{r,s}(\Hat{c}_{p,p'}): \quad   \cdots  F_{r,2p+s}^{N=1}   \overset{(Q_{-}^{F})^{s}}{\longrightarrow}  F_{r,-s +2p j }^{N=1}     \overset{(Q_{-}^{F})^{p-s}}{\longrightarrow}    F_{r,s}^{N=1}  \overset{(Q_{-}^{F})^{s}}{\longrightarrow} F_{r,-s}^{N=1}  \overset{(Q_{-}^{F})^{p-s}}{\longrightarrow}  \cdots 
\end{equation}
We \textbf{conjecture} that irreducible modules in $N=1$ minimal models are isomorphic to the zeroth cohomology space of $C_{r,s}(\Hat{c}_{p,p'})$, with all other cohomology spaces being trivial
\begin{equation}
     H^{i} (C_{r,s}(\Hat{c}_{p,p'})) \cong \begin{cases}
       0 & i \ne 0 ,  \\  \mathcal{H}_{r,s}^{N=1}(\Hat{c}_{p,p'}) , & i= 0. 
    \end{cases} 
\end{equation}

\paragraph{The application to Ishibashi states:}

We apply our conjecture to Ishibashi states in $N=1$ minimal models. From the form of the $T(z)$ and $T_{F}(z)$ of background-charged $N=1$ free field, we write down their modes in terms of the modes of the free fields
\begin{equation}
    L_{n} = \frac{1}{2} \sum_{k \in \mathbb{Z} } : a_{n-k} a_{k}  : + \frac{1}{2}  \sum_{r} (r+\frac{1}{2}) : b_{n-r} b_{r}   : - \alpha_{0}^{N=1} (n+1) a_{n},
\end{equation}
\begin{equation}
    G_{r} =   \sum_{n \in \mathbb{Z} }  (n+1) :  b_{r-n}  a_{n}  :+ 2\alpha_{0}^{N=1} (r+\frac{1}{2} ) b_{r}.  
\end{equation}

The $N=1$ Ishibashi states take the form of 
\begin{equation}
  \vert \alpha, \Bar{\alpha} , \alpha_{0}^{N=1} \rangle \rangle_{N, \pm} =  \exp{ \Big(  \sum_{n=1}^{\infty}  - \frac{\Omega}{n} a_{-n} \Bar{a}_{-n} \Big) } \exp{ \Big(  \pm i \sum_{r=1}^{\infty}  b_{-r} \Bar{b}_{-r}  \Big)  }  \vert \alpha , \Bar{\alpha} ,\alpha_{0}^{N=1}    \rangle .
\end{equation}
Imposing the $N=1$ gluing conditions of NS Ishibashi states $\vert \mathcal{H}_{r,s}^{N=1} (\Hat{c}_{p,p'}) \rangle \rangle_{\pm}$
\begin{equation}
    (L_{n}-\Bar{L}_{-n}) \vert \mathcal{H}_{r,s}^{N=1} (\Hat{c}_{p,p'}) \rangle \rangle_{\pm} =  (G_{r} \pm  i \Bar{G}_{-r}) \vert \mathcal{H}_{r,s}^{N=1} (\Hat{c}_{p,p'}) \rangle \rangle_{\pm} =0 ,  
\end{equation}
to Fock space Ishibashi states. The Virasoro gluing conditions give a similar constrain on the bosonic part
\begin{equation}
   \Omega=1, \quad \alpha-\Bar{\alpha} -2\alpha_{0}^{N=1}=0 .
\end{equation}
From this constraint, we obtain the required bosonic gluing conditions 
\begin{equation}
    (a_{n}-\Bar{a}_{-n}- 2\alpha_{0}^{N=1} \delta_{n,0} ) \vert \alpha, \Bar{\alpha} , \alpha_{0}^{N=1} \rangle \rangle_{N, \pm} =0.
\end{equation}
Combining it with the fermionic gluing condition
\begin{equation}
   ( b_{r}  \mp i   \Bar{b}_{-r} )\vert \alpha, \Bar{\alpha} , \alpha_{0}^{N=1} \rangle \rangle_{\omega, \pm} =0, 
\end{equation}
the $N=1$ condition $(G_{r} \pm  i \Bar{G}_{-r}) \vert \mathcal{H}_{r,s}^{N=1} (\Hat{c}_{p,p'}) \rangle \rangle_{\pm}$ is automatically satisfied. Hence, the $N=1$ free field Ishibashi states involved are $\vert \alpha_{r,s}, 2\alpha_{0}^{N=1} -\alpha_{r,s} , \alpha_{0}^{N=1} \rangle \rangle_{N, \pm}$. 

We write down both the NS and R sectors Ishibashi states as
\begin{equation}
    \vert \mathcal{H}_{r,s}^{N=1} (\Hat{c}_{p,p'}) \rangle \rangle_{\pm} = \sum_{j \in \mathbb{Z} } \kappa_{\pm;j}  \vert  \alpha_{r, \pm s -2jp  },  2\alpha_{0}^{N=1} -\alpha_{r,s} , \alpha_{0}^{N=1}  \rangle  \rangle_{N , \pm}  ,
\end{equation}
where the phases $\kappa_{\pm;j}$ are also undetermined $U(1)$ phases. Similar to the Virasoro Ishibashi state cases, it should satisfy $\kappa'_{\pm;j }\kappa_{\pm;j } = \pm 1 $, where $\kappa'_{\pm;j }$ are undetermined phases in the dual Ishibashi state $_{\pm}\langle \langle \mathcal{H}_{r,s}^{N=1} (\Hat{c}_{p,p'}) \vert $

\

\subsection{Fermionic expressions of Ising model Ishibashi states}

We show the free fermionic expression of the Ishibashi states in the Ising model $\mathcal{M}(4,3)$. The vectors in the irreducible modules of the Ising model are identified with vectors in the free fermion Fock spaces as following
\begin{itemize}
    \item  Vectors in $\mathcal{H}_{1,1}^{4,3}$ are the fermion number even vectors in the NS sector of free fermion Fock space.

    \item  Vectors in $\mathcal{H}_{2,1}^{4,3}$ are the fermion number odd vectors in the NS sector of free fermion Fock space. 

    \item  Vectors in $\mathcal{H}_{1,2}^{4,3}$ can be represented by either fermion number odd or even vectors in the R sector of free fermion Fock space. 
\end{itemize}
The vector identifications lead to the following free fermion expressions of Ising Ishibashi states
\begin{equation}
     \vert \mathcal{H}_{1,1}^{4,3} \rangle \rangle = \frac{1}{2} \big( \vert \psi,  \text{NS} \rangle \rangle_{+}  +   \vert \psi,  \text{NS}  \rangle \rangle_{-} \big) , \quad   \vert \mathcal{H}_{2,1}^{4,3} \rangle \rangle = \frac{1}{2} \big(  \vert \psi, \text{NS}  \rangle \rangle_{+}  -   \vert \psi, \text{NS}  \rangle \rangle_{-} \big) . \label{eq: NS Is fer}
\end{equation}
\begin{equation}
   \vert \mathcal{H}_{1,2}^{4,3} \rangle \rangle =  \frac{1}{2} \big( \vert \psi,  \text{R} \rangle \rangle_{+} \pm \vert \psi,  \text{R}  \rangle \rangle_{-} \big) , \label{eq: R Is fer}
\end{equation}
where the free fermion Ishibashi states are \cite{Recknagel:2013uja} 
\begin{equation}
    \vert \psi, \text{NS}  \rangle \rangle_{\pm} = \exp{\Big[ \pm i \sum_{r\in \mathbb{N} + \frac{1}{2} }  \psi_{-r} \Bar{\psi}_{-r} \Big]} \vert I \otimes \Bar{I} \rangle ,
\end{equation}
\begin{equation}
    \vert \psi, \text{R}  \rangle \rangle_{\pm} = \exp{\Big[ \pm i \sum_{r\in \mathbb{Z}_{+}  }  \psi_{-r} \Bar{\psi}_{-r} \Big]}  (1 \pm i \psi_{0} \Bar{\psi}_{0}) \vert R \otimes \Bar{R} \rangle ,
\end{equation}
where $ \vert R \otimes \Bar{R} \rangle$ is the fermion number even, $(h,\Bar{h})=(\frac{1}{16}, \frac{1}{16})$ R sector ground state. It is easy to check that the overlaps between the fermionic expressions of the Ising Ishibashi states reproduce the correct Ising chiral characters.

\appendix

\acknowledgments

The author is grateful to H. Z. Liang for his guidance and encouragement. The author is grateful to S. Ribault, A. Sagnotti, T. Tada, and J. Vošmera for their valuable comments.




\end{document}